\documentclass[a4paper,10pt]{article}

\usepackage{jheppub} 

\usepackage{graphicx,floatflt,amssymb,epsfig}
\usepackage[utf8]{inputenc}
\usepackage[inline]{enumitem}
\usepackage{mathrsfs}
\usepackage{amsmath}
\usepackage{color}
\usepackage{graphicx}
\usepackage{subfig}
\usepackage{multirow}
\usepackage{mathtools}
\usepackage{float}
\usepackage{pstricks}
\usepackage[toc,page]{appendix}
\usepackage{pifont}
\usepackage{braket}
\usepackage{ragged2e}
\usepackage{slashed}
\usepackage{rotating}
\usepackage{tablefootnote}

\usepackage{hyperref}
\usepackage{tikz}
\usepackage{tikz-feynman}
\usetikzlibrary{arrows}
\usepackage{caption} 
\renewcommand{\arraystretch}{1.5}

{\newcommand{\lsim}{\mbox{\raisebox{-.6ex}{~$\stackrel{<}{\sim}$~}}}
	{

		\newcommand{\JP}{{J/\psi}}

		\newcommand{\bqa}{\begin{eqnarray}}
		\newcommand{\eqa}{\end{eqnarray}}
		\newcommand{\stateinequ}[3]{{}^{#1}{#2}_{#3}}

		\newcommand{\bmt}{\begin{pmatrix}}
			\newcommand{\emt}{\end{pmatrix}}
		\newcommand{\ba}{\begin{array}{c}}
			\newcommand{\ea}{\end{array}}
		\newcommand{\beq}{\begin{equation}}
		\newcommand{\eeq}{\end{equation}}
		\newcommand{\bea}{\begin{eqnarray}}
		\newcommand{\eea}{\end{eqnarray}}
		\newcommand{\nn}{\nonumber}
		\newcommand{\bi}{\begin{itemize}}
			\newcommand{\ei}{\end{itemize}}
		
		\newcommand{\baz}{\begin{array}{cc}}
			
			\newcommand{\mathsym}[1]{{}}

			\newcommand{\bt}{\begin{tabular}}
				\newcommand{\et}{\end{tabular}}

			\newcommand{\benu}{\begin{enumerate}}
				\newcommand{\eenu}{\end{enumerate}}
			
			\newcommand{\bav}{\begin{array}{cccc}}

				\allowdisplaybreaks

	\title{\boldmath Analyzing the semileptonic and nonleptonic $B_c \to J/\psi, \eta_c$ decays   }
	
	\author[a]{Aritra Biswas}
	\emailAdd{iluvnpur@gmail.com}
	
	\author[b]{Soumitra Nandi}
	\emailAdd{soumitra.nandi@iitg.ac.in}
	
	\author[b]{Shantanu Sahoo}
	\emailAdd{shantanu\_sahoo@iitg.ac.in}
	
	\affiliation[a]{Universitat Aut\`onoma de Barcelona, 08193 Bellaterra, Barcelona,\\
		Institut de F\'{i}sica d'Altes Energies (IFAE), The Barcelona Institute of Science and Technology, Campus UAB, 08193 Bellaterra (Barcelona) }	
	\affiliation[b]{Indian Institute of Technology Guwahati, North Guwahati, Assam 781039, India}

\abstract{
This study focuses on the decay of the $B_c$ meson to S-wave charmonia. Starting with lattice inputs on $B_c\to J/\psi$ form factors, we have obtained the $B_c\to\eta_c$ form factors using heavy quark spin symmetry (HQSS) relations between the associated form factors after parametrizing and extracting the possible symmetry breaking corrections. Using the $q^2$ shapes of these form factors, we have computed the branching fractions $\mathcal{B}(B_c^-\to \eta_c\ell^-\bar{\nu})$ (with $\ell =\tau, \mu (e)$) and the decay rate distributions and have predicted the Standard model estimate for the observable $R(\eta_c)=\Gamma(B_c^-\to \eta_c\tau^-\bar{\nu})/\Gamma(B_c^-\to \eta_c\mu^-\bar{\nu}) =0.310 \pm 0.042$. In addition, we have estimated the radial wave functions $\psi_{B_c}^R(0)$, $\psi_{J/\psi}^R(0)$ and $\psi_{\eta_c}^R(0)$ at small quark-antiquark distances from the available information on the form factors from lattice, the lattice inputs on the decay constants $f_{J/\psi}$, $f_{\eta_c}$, $f_{B_c}$ and experimental data on radiative and rare decays of the $J/\psi$ and $\eta_c$ mesons. To do so, we choose the theory framework of non-relativistic QCD (NRQCD) effective theory. Using our results, we have predicted the branching fractions of a few non-leptonic decays of $B_c \to J/\psi$ or $\eta_c$ and other light mesons. We have also updated the numerical estimates of the cross sections $\sigma(e^+e^- \to J/\psi \eta_c, \eta_c\gamma)$ and predicted the branching fractions of $Z$ boson decays to either $J/\psi$ or $\eta_c$ final states or both.}

\arxivnumber{}

\begin{document}
	\maketitle
\section{Introduction}

In the recent past and beyond, experimental collaborations like the B-factories and LHCb have gathered a plethora of data on the various decays of $B_u$, $B_d$ and $B_s$ mesons which have helped in improving our understanding of the underlying low-energy Quantum Chromodynamics(QCD). The remarkable developments in lattice calculations have also been very useful. With improved precision, data might help pinpoint any underlying beyond the standard model (BSM)  dynamics.    

The semileptonic $b\to c\ell^-\bar{\nu}$ (with $\ell = e, \mu, \tau$) decays have gained a lot of attention over the past decade both from theorists as well as from experimental collaborations like Belle and LHCb. The modes with light leptons are used for the extraction of the Cabibbo-Kobayashi-Maskawa (CKM) matrix element $V_{cb}$ while the mode with $\ell =\tau$ is expected to help probe new physics (NP) beyond the standard model (SM). Observables like ratios of the decay rates $R(D^{(*)}) = \Gamma(\bar{B}\to D^{(*)}\tau^- \bar{\nu})/\Gamma(\bar{B}\to D^{(*)} \mu^-(e^-) \bar{\nu} )$ have been predicted in the SM and measured by experimental collaborations; see \cite{ParticleDataGroup:2020ssz,HFLAV2019} for a review. At the present level of precision, data shows a combined deviation of $\sim 3\sigma$ from SM estimates~\cite{HFLAV2019}, allowing for sizeable NP contributions to these modes \cite{Ray:2023xjn, Fedele:2023ewe}. NP analyses based on relatively old data can be seen, for example, in  refs. \cite{Azatov:2018knx, Bhattacharya:2018kig, PhysRevD.98.095018, Angelescu:2018tyl, Iguro:2018fni, Murgui:2019czp, Shi:2019gxi, Becirevic:2019tpx, Biswas:2021pic, Iguro:2022yzr}. For a model independent simultaneous explanation of the data on $R(D)$ and $R(D^*)$, contribution from a new SM type four-fermion operator is required assuming contribution from a single operator at a time, see for example the result presented in table 7 of ref.\cite{Ray:2023xjn}. Other one-operator scenarios, consisting of right-handed vectorial, scalar, pseudoscalar or tensor type four-fermion operators have difficulty in explaining the data simultaneously. However, contributions from two operators at a time can explain the data simultaneously in almost all the scenarios with vector, axial-vector, scalar-pseudoscalar and tensor type operators; see the results of table 9 of ref. \cite{Ray:2023xjn}. For a more robust conclusion, we need more precise inputs from lattice and experimental collaborations.     


In order to gain complementary phenomenological information compared to those mentioned above concerning well-analysed mesonic decays, the study of various other similar decay modes is essential. Such studies can be helpful in improving our understanding of the nature of the anomalous results seen in B-meson decays. Moreover, any BSM physics altering the results for these modes should affect and be constrained by other $b \to c$ transitions. Among all such processes, the decay of the $B_c$ meson is of considerable interest. In comparison to the $B_u$, $B_d$ and $B_s$ mesons, $B_c$ has some special properties:
\begin{itemize}
	\item $B_c$-meson is a heavy quarkonium with mixed-heavy flavours: the bottom (b) and the charm (c) quarks. The decays of $B_c$ meson could be via the weak decays of its $b$ and/or $\bar{c}$ quark. Hence, the weak decays of $B_c$ are instrumental for comparative studies of the two heavy flavours and provide unique opportunities to study heavy-quark dynamics. For example, the $B_c$ can be used in estimating and measuring the ratio of CKM matrix elements $\frac{|V_{cb}|}{|V_{cs}|}$, where cancellation of the theoretical uncertainties as well as the experimental systematic errors in the $B_c$ meson decay takes place. In addition, the $(b\bar{c})$ system is interesting since it allows one to use the phenomenological information obtained from a detailed experimental study of the charmonium and bottomonium.
	
	\item LHCb is expected to produce around $5 \times 10^{10}$ $B_c$ events per year~\cite{Gouz:2002kk, yuan2010experimental}, and one can hope to gather insights about the decay of the $B_c$ meson with a degree of rigour that was hitherto not possible. 
	
	\item The lightest pseudoscalar $B_c$ meson is stable against strong and electromagnetic interactions as it lies below the $B\bar{D}$ threshold. It can only decay through weak interactions with an average lifetime $\tau_{B_c}=0.510(9) \text{ps}$~\cite{ParticleDataGroup:2022pth}, making it an ideal system to study weak decays of heavy quarks. 
	
	\item Since both of its constituents are heavy, each of these can decay individually with the other as a spectator. This possibility offers a promising opportunity to study various non-leptonic and semileptonic weak decays of heavy mesons, which is helpful to test the SM and reveal any BSM physics. 
\end{itemize}

The decay mode of the $B_c$ meson involving a $J/\psi$ can be relatively easily reconstructed. This is because $J/\psi$ can further decay into a di-lepton pair $\ell^+\ell^-$ ($\ell=\mu^-, e^-$) with a probability of 10\%~\cite{ParticleDataGroup:2022pth}. Leptonic channels are experimentally cleaner in comparison to channels with hadronic final states that suffer from gluon contamination and can hence be used for more efficient reconstruction of the initial state particle. Specifically, LHCb has a greater efficiency of muon channel extraction ($\sim 95-98\%$ ref.~\cite{Archilli:2013npa}) and the probability of hadrons being misidentified as muons is at the level of $1\%$. Compared to $J/\psi$, the $\eta_c$ decays mostly to hadronic final states. The probability for an $\eta_c\to\gamma\gamma$ transition (the only non-hadronic final state that $\eta_c$ decays to according to~\cite{ParticleDataGroup:2022pth}) is 0.01\%. The CDF collaboration ref.~\cite{CDF:1998ihx} first discovered the $B_c$ meson via the $B_c^- \to J/\psi \ell^-\bar{\nu}$ decays. In the near future, this semileptonic decay can be useful to extract $|V_{cb}|$. Furthermore, one can define a ratio $R(J/\psi)$ similar to the one as defined above for probing NP. The NP effects in $\bar{B} \to D^{*}\tau^- \bar{\nu}$ decays will be highly correlated with those in $ B_c^- \to J/\psi \tau^-\bar{\nu}$ \cite{Bhattacharya:2018kig}. The LHCb collaboration has measured this ratio to be~\cite{LHCb:2017vlu}:
\begin{equation}
R(J/\psi) = \frac{\Gamma(B_c^- \to J/\psi \tau^-\bar{\nu})}{\Gamma(B_c^- \to J/\psi \mu^-\bar{\nu})} = 0.71 \pm 0.17 (stat) \pm 0.18 (syst)= 0.71 \pm 0.25.
\label{eq:Rjpsilhcb}
\end{equation}
As has been stated in ref.~\cite{Cohen:2018dgz}, the $95\%$ confidence level $0.20 \le R(J/\psi) \le 0.39$ for the SM estimate of $R_{J/\psi}$ agrees with the LHCb result at the $1.3\sigma$ level.
The CMS collaboration has also measured the same ratio and quote~\cite{CMS-PAS-BPH-22-012}:
\begin{equation}
R(J/\psi)= 0.17^{+0.18}_{-0.17} (stat.)^{+0.21}_{-0.22} (syst.)^{+0.19}_{-0.18} (theo)= 0.17\pm 0.33.
\label{eq:CMSres}
\end{equation}
As will be discussed in the following section, several QCD models exist in the literature from a theoretical perspective. Based on the modelling of the form factors, the value for the SM prediction of $R(J/\psi)$ lies in the range $[0,0.48]$~\cite{Anisimov:1998uk, Kiselev:1999sc, Kiselev:2002vz, Ivanov:2006ni,Hernandez:2006gt, Qiao:2012vt,Wang:2012lrc,Rui:2016opu,Dutta:2017xmj, Watanabe:2017mip, Issadykov:2018myx,Leljak:2019eyw,Cohen:2018dgz}. A model-independent approach regarding the form factors~\cite{Cohen:2019zev} leads to the SM prediction of $0.25(3)$. The HPQCD lattice collaboration has recently extracted the $B_c \to J/\psi$ form-factors over the full kinematically allowed region \cite{Harrison:2020gvo}. They have predicted $R(J/\psi) = 0.2582 (38)$ \cite{Harrison:2020nrv}, which is so far the most precise prediction and in tension at $ \sim 1.8 \sigma $ level with the LHCb result given in eq.~\ref{eq:Rjpsilhcb}. The CMS result agrees with the SM predicted value at 0.3$\sigma$ and  with the LHCb measurement at 1.3$\sigma$; whereas the experimental average, $R(J/\psi)^{\text{avg.}}_{\text{expt.}}$=$\text{0.51 $\pm$ 0.20}$ (weighted average) is compatible with the SM at $\text{1.3$\sigma$}$.  

Another equally important decay mode is $B_c^- \to \eta_c \ell^-\bar{\nu}$ which is analogous to the $\bar{B}\to D\ell^-\bar{\nu}$ decay. So far, we don't have sufficient input on the form factors relevant for this decay from lattice, nor do we have any measurement available. In this paper, we have extracted the information on $B_c\to \eta_c$ form-factors using the available information on $B_c \to J/\psi$ form-factors from lattice in combination with the heavy-quark-spin-symmetry (HQSS), and thus predicted the observable $R({\eta_c})$. In the non-relativistic QCD (NRQCD) effective theory framework, the $B_c \to J/\psi$ and $B_c\to \eta_c$ form-factors have been calculated including the next-to-leading order corrections at $\mathcal{O}(\alpha_s^2)$ \cite{Bell:2006tz, Qiao:2012vt, Qiao:2012hp} and the relativistic corrections \cite{Zhu:2017lqu}. The numerical evaluation of these form factors requires knowledge of the model-dependent non-perturbative matrix elements, which are essentially the respective wave functions at the origin of the charmonium and the $B_c$ meson. We have extracted these matrix elements from the available information on the $B_c \to J/\psi$ form-factors from lattice. We have also used the measured rates of the leptonic and radiative decays of $J/\psi$ and $\eta_c$ in the fit to extract the matrix elements. 

Finally, we have predicted $R(\eta_c)$ using the fit results of the associated non-perturbative matrix elements relevant for $B_c$ meson and the charmoniums $J/\psi$ and $\eta_c$. We have updated several other predictions related to $e^+e^-$, Z-decays to single or double charmonium and radiative modes involving $J/\psi$ and/or $\eta_c$. Furthermore, we have updated the predictions for the branching fractions of a few non-leptonic decay modes of the $B_c$, including a light meson and $J/\psi$ or $\eta_c$ in the final state.

\section{Study of $B_c^-\to\eta_c\ell^-\bar{\nu}_{\ell}$ decay: Form factors and the ratio $R(\eta_c)$}\label{sec:Bctoetac}

In order to predict the ratio of the decay rates $R(\eta_c)$, we need a proper knowledge of the $q^2$ distributions of the $B_c^-\to\eta_c\ell^-\bar{\nu}$ decays. To obtain an  estimate for the shape of the differential decay width distribution over the entire allowed kinematical range, knowledge of the behaviour of the non-perturbative aspects of the decay, i.e. the form factors over the whole di-lepton range is imperative. As of now, we do not have any reliable lattice estimates on the $B_c\to\eta_c$ form factors $f_{+,0}^{B_c\to\eta_c}$ from lattice collaborations. Ref.~\cite{Colquhoun:2016osw} provides certain estimates, albeit with an incomplete error treatment. We hence refrain from using this input in our analysis.

Note that the lattice inputs on the $B_c\to J/\psi$ form-factors are available in the full kinematically allowed range of $q^2$ \cite{Harrison:2020gvo} which could be instrumental in constraining the $q^2$ behaviour of the decay rate.
In ref.~\cite{Harrison:2020gvo}, a fit to the simulated data has been carried out following a specialized parametrization for the form factors defined over the entire di-lepton mass invariant region for the corresponding form factors. The details of the relevant inputs used in this parametrization can be seen from \cite{Harrison:2020gvo} which we have also incorporated in our analysis. The authors provide complete information regarding their fits. Hence, using the central values, uncertainties and correlations among the coefficients of the polynomial series provided by lattice, we generate synthetic data at maximum recoil ($q^2=0$) for our analysis. These data points are shown in table \ref{tab:inputlattQCDSR}. We use these inputs on $B_c\to J/\psi$ form factors and theoretical tools like heavy quark spin symmetry (HQSS)~\cite{Jenkins:1992nb, Kiselev:1999sc} to predict the form factors $f_{+,0}^{B_c\to\eta_c}$. In this section, we will discuss the details of the methodology used to constrain the shape of $f_{+,0}^{B_c\to\eta_c}(q^2)$. Following heavy quark effective theory (HQET), all transition matrix elements between two hadrons with a single heavy-quark (anti-quark) along with a light anti-quark (quark) are related via a universal function called the ``Isgur-wise" function. However, this approximation fails for a heavy-heavy bound-state system (as is the case for $B_c$ decays). This is due to the fact that both of the constituent quarks are now of comparable masses. This results in a broken Heavy-Quark Flavor Symmetry. However, the decaying and the spectator quarks still retain their separate heavy-quark spin symmetries. And in fact, such systems are described to a better degree of accuracy by non-relativistic dynamics since the valence quarks are heavy. Therefore, the six transition form factors for $B_c\to J/\psi$ and $B_c\to\eta_{c}$ can be related to a single universal function $\Delta$ via the spin symmetries (instead of the flavour symmetry for the case of the ``Isgur-Wise" functions) \cite{Jenkins:1992nb,Kiselev:1999sc} which may receive symmetry breaking correction $\mathcal{O}(m_c/m_b,  \Lambda_{QCD}/m_b) \approx 30\%$, for detail see the refs.\cite{Jenkins:1992nb, Cohen:2019zev}. The trace formalism developed in \cite{Falk:1990yz} is useful to find the relative normalization between these six form factors using NRQCD near the zero recoils \cite{Jenkins:1992nb,Kiselev:1999sc} and at the non-zero recoils \cite{Cohen:2019zev}. Here, the recoil variable $w$ is defined as $w = (m_{B_c}^2+m_{F}^2-q^2)/(2m_{B_c}m_{F})$, where $m_F$ is the final state meson.

From ref.~\cite{Cohen:2019zev}, following the leading order NRQCD relations between the form factors in $B_c\to J/\psi$ and $B_c\to \eta_c$, we obtain
\begin{eqnarray}
V &=& - \frac{m_{B_c}+m_{J/\psi}}{2}  \frac{c_V}{m_{B_c}\sqrt{r}}\Delta_V(w) \, , \nonumber\\
A_1 &=& \frac{m_{B_c}}{(m_{B_c}+m_{J/\psi})}\sqrt{r}c_\epsilon\Delta_{A_1}(w)  \, , \nonumber\\
A_2 &=& -\frac{(m_{B_c}+m_{J/\psi})}{2 m_{J/\psi}}\sqrt{r} (r c_1 +c_2) \Delta_{A_2}, \nonumber \\
A_0 &=& \frac{c_\epsilon+ (1-wr)c_1 + (w-r)c_2} {2 \sqrt{r}}
\Delta_{\mathcal{A}_0}(w) \, ,\nonumber\\
f_{+}&=&\sqrt{r} \frac{c_1^P + c_2^P r^{-1}}{2}\Delta_+(w) \, ,\nonumber\\
f_0&=&\frac{1}{(m_{B_c}^2-m_{\eta_c}^2)}m_{B_c}^2 \sqrt{r}\big[(1-wr)c_1^P +(w-r)c_2^P\big]\Delta_0(w) \,  . \label{eqn:FFrelns}
\end{eqnarray}
Here, we have parametrised these $\Delta_i(w)$ functions as given below following a Taylor series expansion around $w=1$,
\begin{equation}
\Delta_i(w)= \Delta_i(1)+(w-1)\Delta(1)'+\frac{(w-1)^2}{2}\Delta(1)''.
\label{eqn:Delta}
\end{equation}
The unknown parameters $\Delta_i(1)$, $\Delta(1)'$ and $\Delta(1)''$ can be estimated via a $\chi^2$ optimization method using lattice results on $B_c \to J/\psi$ form factors. Note that here the parameters $\Delta(1)'$ and $\Delta(1)''$ are useful in determining the slope of the shapes of the form factors. The definitions for the r, $c_i$'s, $\theta$ and $\omega$ parameters are obtained from ref.~\cite{Cohen:2019zev} and the references therein and are summarized in appendix~\ref{eq:hqsscoeff}. According to HQSS near the zero recoil ($w=1$):
\begin{equation}\label{eq:hqssw1}
\Delta_V(1) = \Delta_{A_1}(1) = \Delta_{A_2}(1) = \Delta_{A_0}(1) = \Delta_+(1) = \Delta_0(1) = \Delta.
\end{equation}
We have mentioned earlier that there could be a large symmetry-breaking correction to these relations, which should be taken into account in case one is willing to use these exact HQSS symmetry relations at $w=1$. In our analysis, we have considered $\Delta_V(1)$, $\Delta_{A_1}(1)$, $\Delta_{A_2}(1)$ and $\Delta_{A_0}(1)$ as independent parameters and fitted them from the lattice inputs\footnote{Alternatively, one can use the HQSS relation given in eq. \ref{eq:hqssw1} and fit $\Delta$ from lattice inputs. However, in such a case, to take into account the symmetry-breaking corrections in the limit $w \to 1$, one should do the following replacements $\Delta_i(1)\to \Delta (1+\delta^i_{\text{corr.}})$. In the fit, the correction term $\delta^i_{\text{corr.}}$ can be treated as nuisance parameters which follow a Gaussian distribution with mean value 0 and 1$\sigma$ uncertainties 0.3. We have checked that both methods will give identical results for $\Delta_{i}(1)$s. This is not surprising since our inputs are the same for both methods.}. As we will see later, to extract $\Delta_+(1)$ and $\Delta_0(1)$, we have utilised the HQSS symmetry and have taken into account the error due to the symmetry breaking corrections. For simplicity and lack of a sufficient number of inputs, we require $\Delta(1)'$ and $\Delta(1)''$ to be the same for all the form factors. As we shall see shortly, the fitted values of $\Delta_i$'s are all consistent with each other within their $1\sigma$ confidence intervals (CI). Note that the form factors defined in eq. \ref{eqn:FFrelns} are dependent on the charm quark mass $m_c$, and the charm quark mass is scheme-dependent. In our analysis, we have used $m_c=1.34\pm 0.27$, which we have obtained from the average of its values in the kinetic \cite{PhysRevLett.114.061802}, pole-mass \cite{PhysRevLett.114.061802} and $\overline{MS}$ \cite{ParticleDataGroup:2020ssz} schemes.

\begin{table*}[t!]
\begin{center}
 \resizebox{0.75\textwidth}{!}{
\begin{tabular}{|*{3}{c|}}
\hline
\multicolumn{1}{|c|}{HQSS parameters} & {Fit results (inputs: $V, A_0,A_1, A_2$) }  &{Fit results (inputs: $V, A_0,A_1$) } \\

\hline\hline
$\text{$\Delta _{V} $(1)}$  &  $0.816(41)$  &   $0.818(41)$  \\
$\text{$\Delta_{A_1}$(1)}$ & $0.842(27)$&   $0.842(27)$  \\
$\text{$\Delta_{A_0}$(1)}$  &  $0.848(28)$ &  $0.865(31)$  \\
$\text{$\Delta_{A_2}$(1)}$  &  $0.836(57)$    & - \\
$\text{$\Delta (1)'$}$  &  $-2.057(89)$ &  $-2.111(93)$  \\
$\text{$\Delta (1)''$}$  &  $4.86(45)$  & $5.21(48)$ \\
\hline
\end{tabular}
  }
\caption{Results of a $\chi^2$-optimization fit for the HQSS parameters. The fit results shown in the second column are obtained using the lattice inputs on $V, A_0, A_1,$ and $A_2$. The results of the third column are obtained after dropping the lattice input on $A_2$.}
\label{tab:HQSSfitparam}
\end{center}
\end{table*}

\begin{figure*}[ht!]
	\centering
	\includegraphics[scale=0.35]{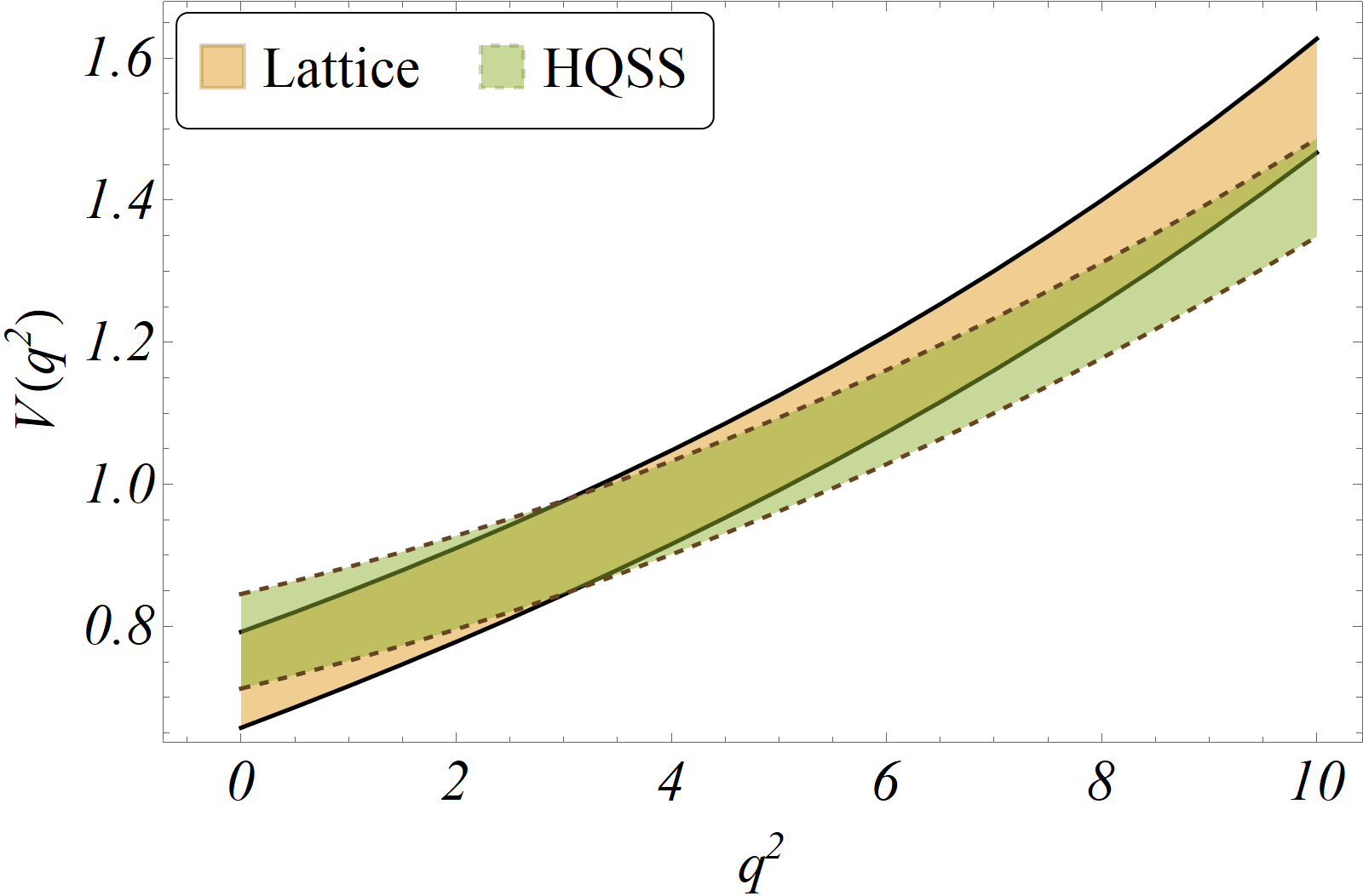}~~~
	\includegraphics[scale=0.35]{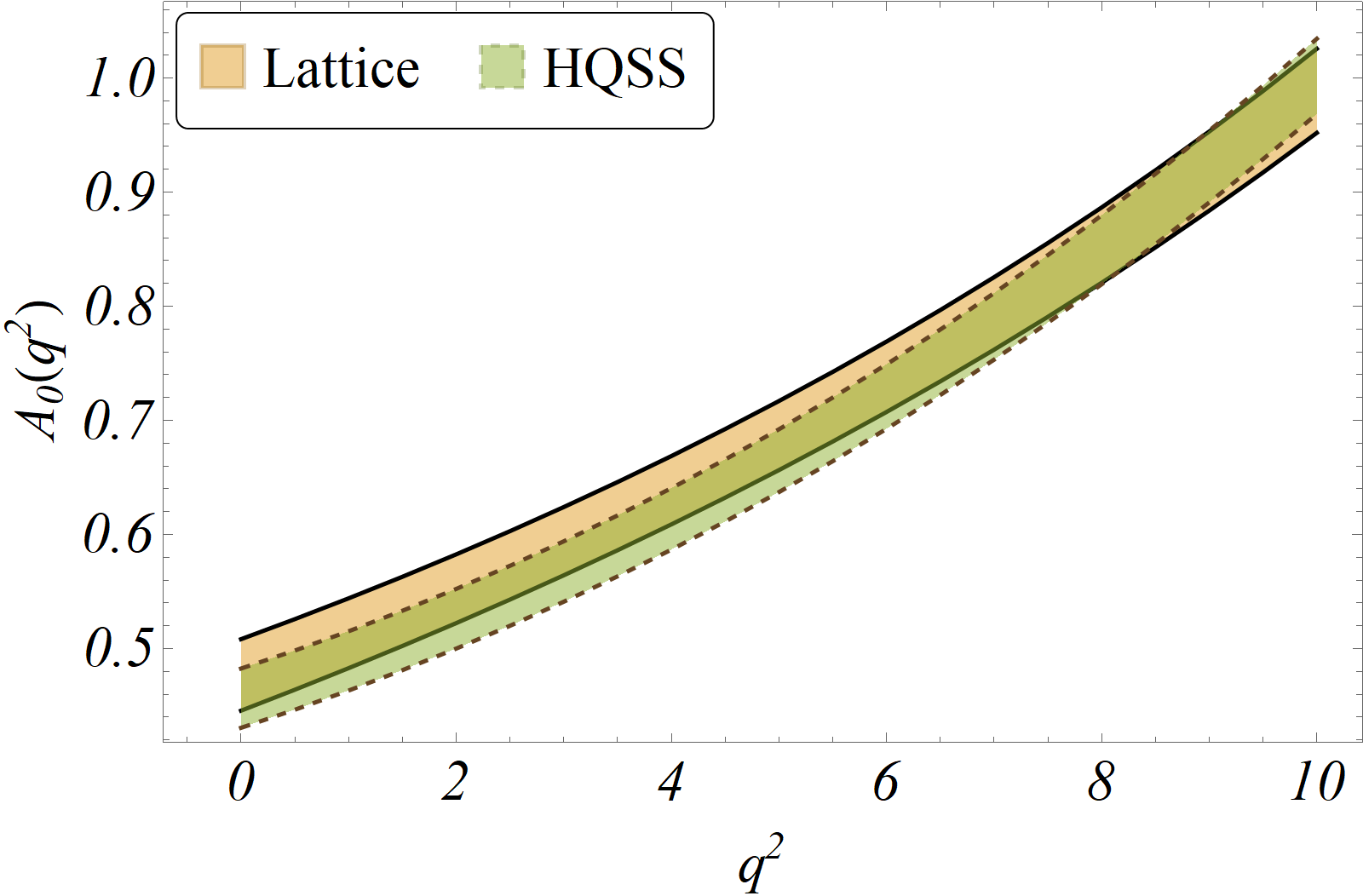}\\
	\includegraphics[scale=0.35]{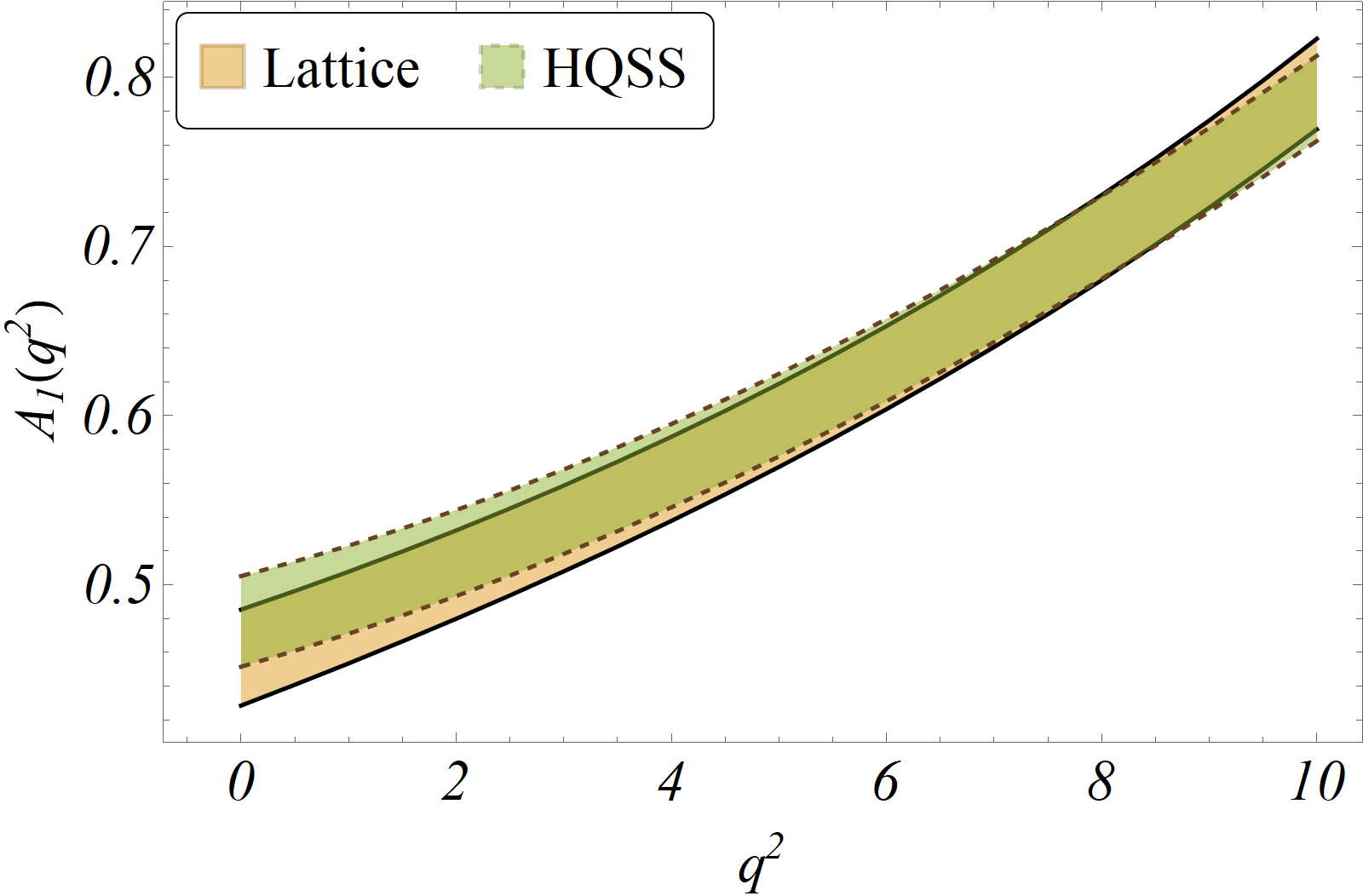}~~~
	\includegraphics[scale=0.35]{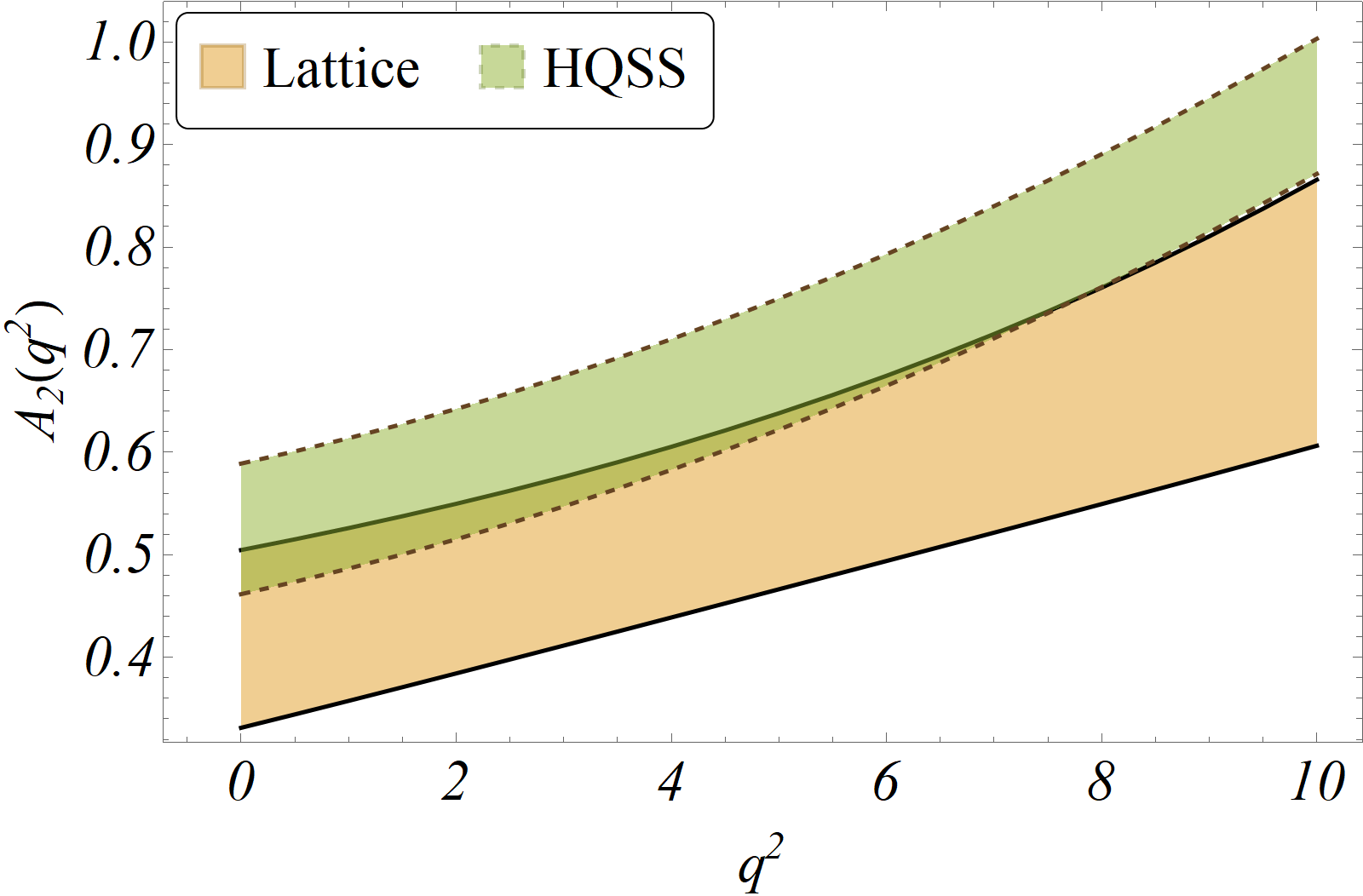}
	\caption{The dashed green bands represent the $q^2$ shapes of the $B_c\to J/\psi$ form factors obtained using the fit results of the HQSS parameters given in the second column of table \ref{tab:HQSSfitparam}. The respective 1$\sigma$ estimates are compared with the corresponding estimates from Lattice (solid brown bands).}
	\label{fig:Bc2psifffit1}
\end{figure*}

We generate synthetic data points for the form factors in $B_c\to J/\psi$ decays over the allowed kinematical range using the fit results and the correlations for the same given in ref.~\cite{Harrison:2020gvo}. Using the generated pseudo data points for all the form factors: $V$, $A_0$, $A_1$ and $A_2$ and their correlations, which are shown in the appendix (table~\ref{tab:corrBcjinputs}), we carry out the fit using the $\chi^2$-minimization procedure to extract the parameters $\Delta_i(1)$, $\Delta(1)'$ and $\Delta(1)''$, respectively. The $\chi^2$ function is defined as\footnote{All the fits and subsequent analyses have been carried out using a \textit{Mathematica}\textsuperscript{\textcopyright} package\cite{OptEx}.}:
\begin{equation}\label{eqn:chi2}
\chi^2=\sum_{i,j}(O_i^{HQSS}-O_i^{lattice})^T V^{-1}_{ij} (O_j^{HQSS}-O_j^{lattice}).
\end{equation}
 Here, $(O^{HQSS}-O^{lattice})$ is a column vector whose $i^{th}$ element is the difference between the theoretical expression and the respective lattice inputs for the $i^{th}$ observable. $V_{ij}$ is the covariance matrix, the $i^{th}$ diagonal element of which is the squared uncertainty corresponding to the lattice estimate for the $i^{th}$ observable. The $\chi^2$ function depends on the concerned parameters through the $O_i^{HQSS}$'s. We note that all the $\Delta_i$'s are consistent with each other within their $1\sigma$ CI's. The corresponding $q^2$ shapes of the form factors obtained from the fit results are shown in fig.~\ref{fig:Bc2psifffit1}. These shapes have been compared to the respective lattice results. The fit results based on HQSS symmetry can correctly reproduce the slopes for the respective form factors and the corresponding $q^2$ distributions for $V(q^2)$, $A_0(q^2)$, and $A_1(q^2)$. Furthermore, we note that we can correctly reproduce the slope for $A_2(q^2)$. However, the numerical values at different $q^2$ points are only marginally consistent with the respective lattice results at 1$\sigma$ CI.
\begin{figure*}[ht!]
	\centering
	\includegraphics[scale=0.35]{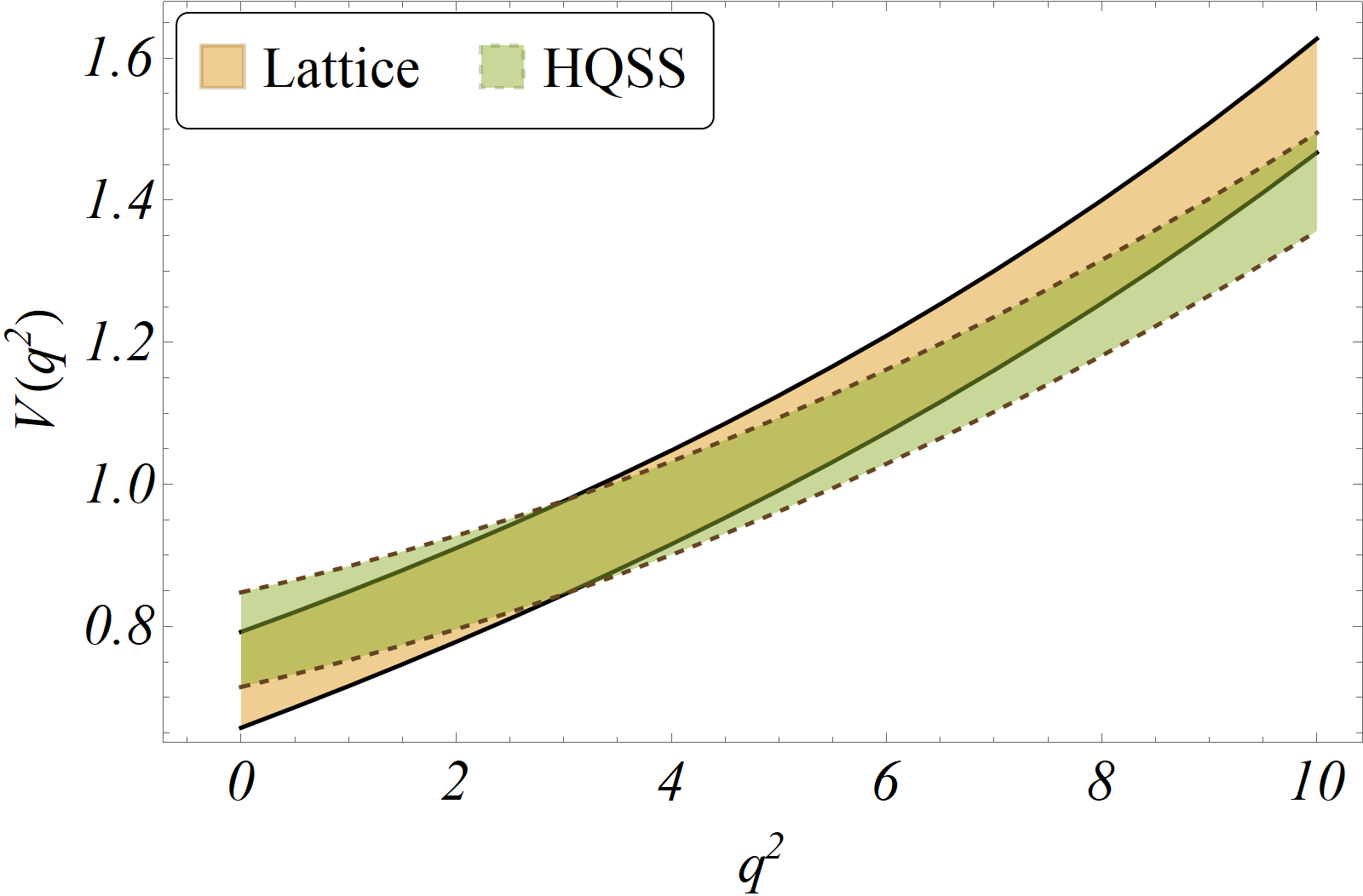}~~~
	\includegraphics[scale=0.35]{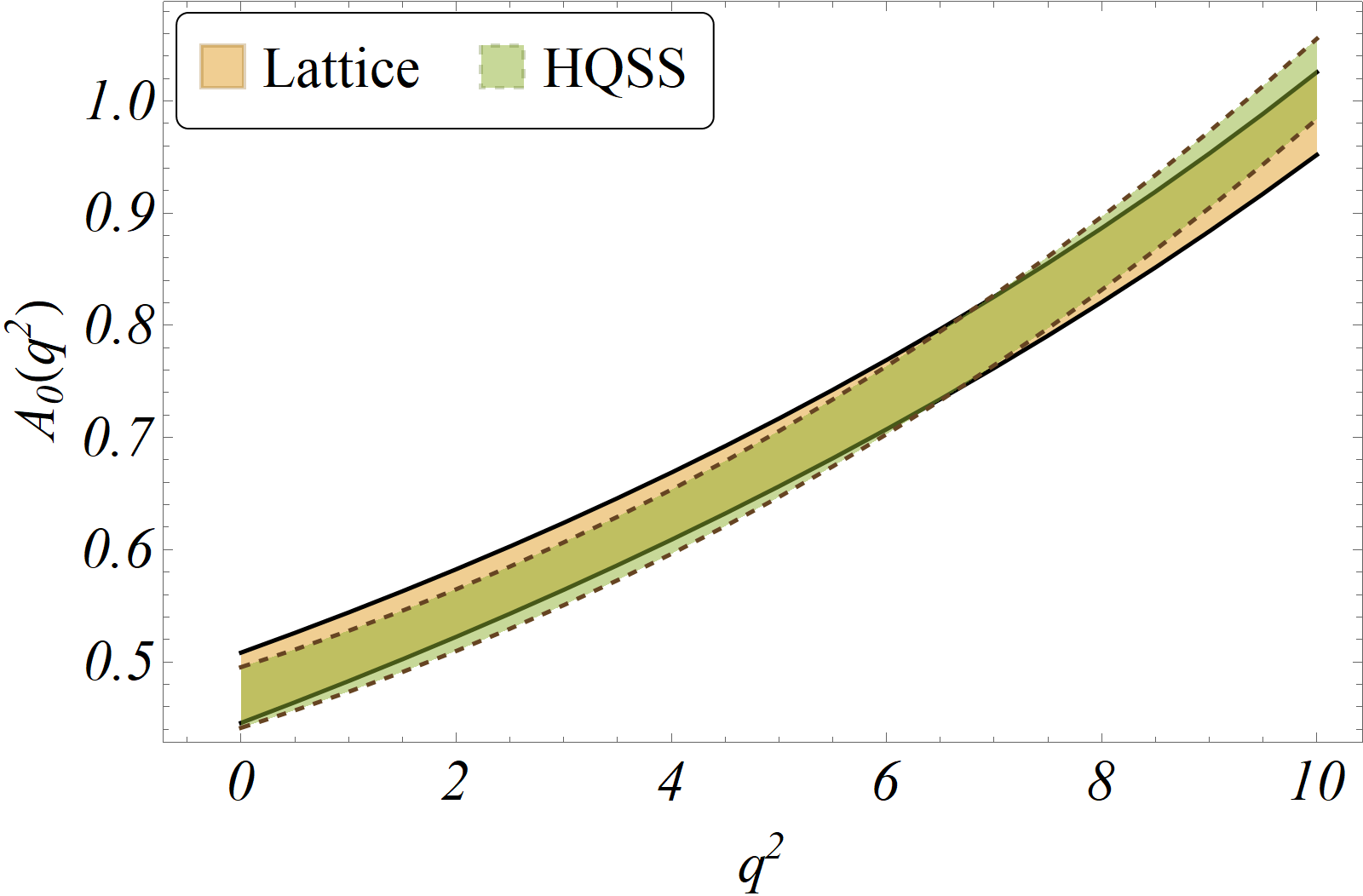}\\
	\includegraphics[scale=0.35]{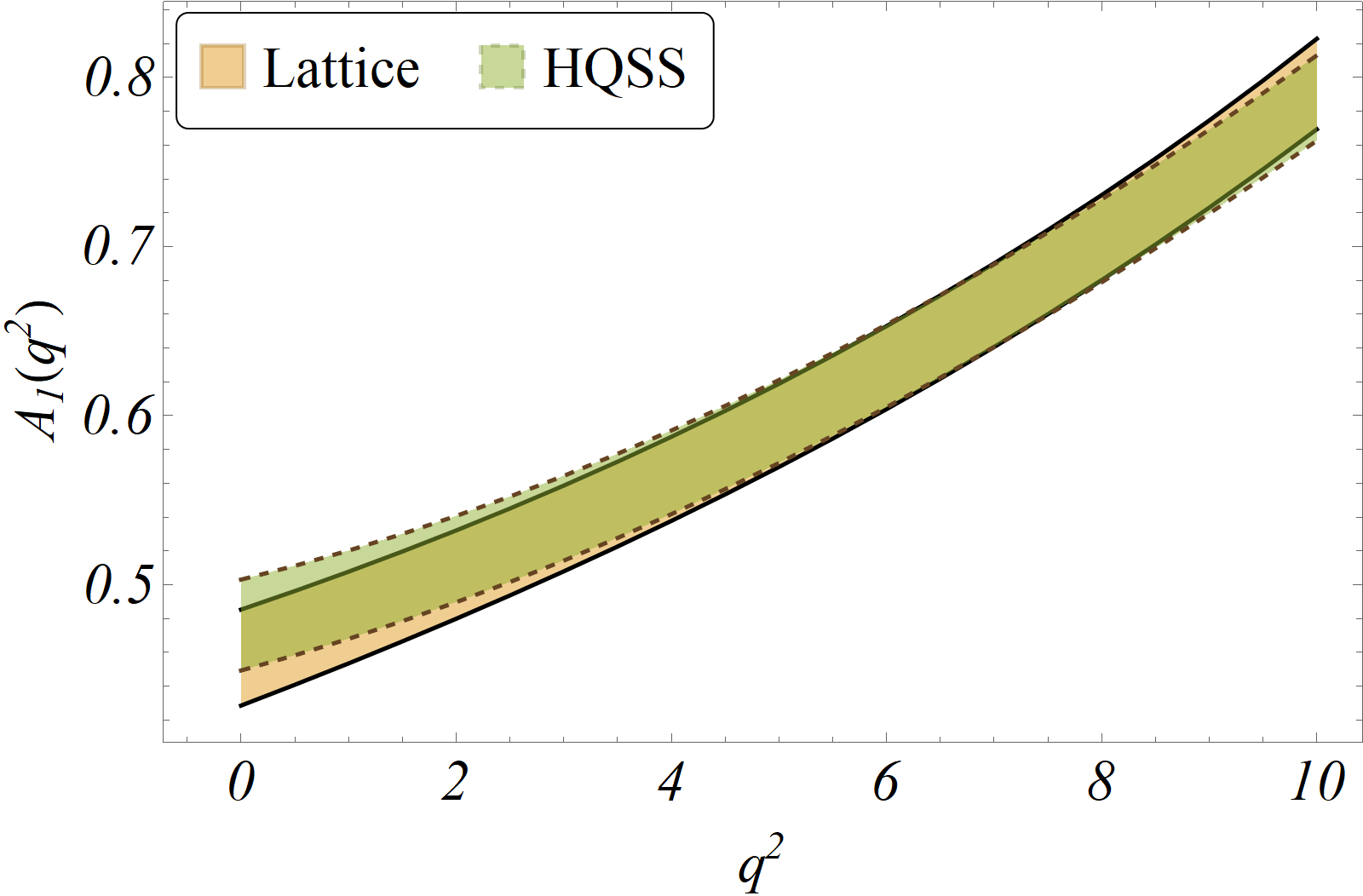}~~~
	\includegraphics[scale=0.35]{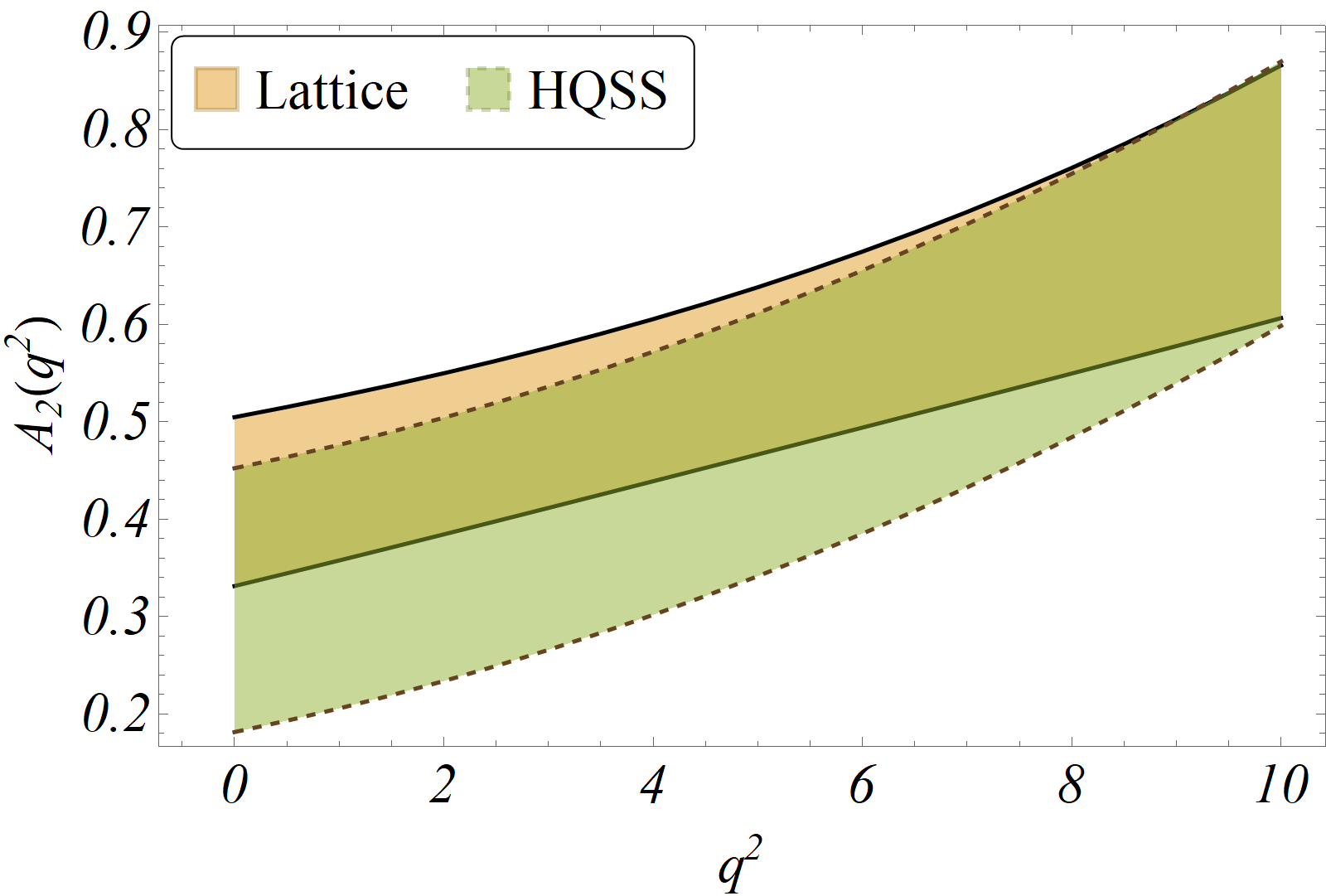}
	\caption{The dashed green bands represent the $q^2$ shapes of the $B_c\to J/\psi$ form factors obtained using the fit results of the HQSS parameters given in the third column of table \ref{tab:HQSSfitparam}. The respective 1$\sigma$ estimates are compared with the corresponding estimates from Lattice (solid brown bands).}
	\label{fig:Bc2psifffit2}
\end{figure*}
Following the above observation, we have carried out another fit where we have excluded the inputs on $A_2(q^2)$. The corresponding fit result is shown in the third column of table~\ref{tab:HQSSfitparam}.
For this fit, we have presented the respective correlation matrix in table \ref{tab:corrHQSSparm2}. We find the results to be almost identical to the one obtained before in the fit with the pseudo data points on $A_2(q^2)$. Using these results we can obtain the shape of $A_0(q^2)$, $A_1(q^2)$ and $V(q^2)$ respectively, which are shown in figure~\ref{fig:Bc2psifffit2}. To obtain the shape of $A_2(q^2)$, we need to fix $\Delta_{A_2}(1)$ (see eqn.~\ref{eqn:Delta}), which we obtain directly by solving the eqn.~$A_2(w=1)=0.74(13)$ to obtain:
\begin{equation}\label{eq:deltaA2}
\Delta_{A_2}(1)=0.65\pm 0.12.
\end{equation}
Using this result for $\Delta_{A_2}(1)$, $\Delta(1)'$ and $\Delta(1)''$ from  table~\ref{tab:HQSSfitparam} ($3^{rd}$ column), we have obtained the shape of $A_2(q^2)$ which has been shown in the bottom right plot of figure \ref{fig:Bc2psifffit2}\footnote{As we missed the correlation information between $\Delta_{A_2}(1)$ with $\Delta(1)^\prime$ and $\Delta(1)^{\prime\prime}$, our form factor 1$\sigma$ error band for $A_2(q^2)$ is larger than the existing lattice prediction.}. Thus, we are able to reproduce the $q^2$ shapes of all four form factors as predicted by lattice from our fits.

\begin{figure*}[t]
	\centering
	\subfloat[]{\includegraphics[scale=0.35]{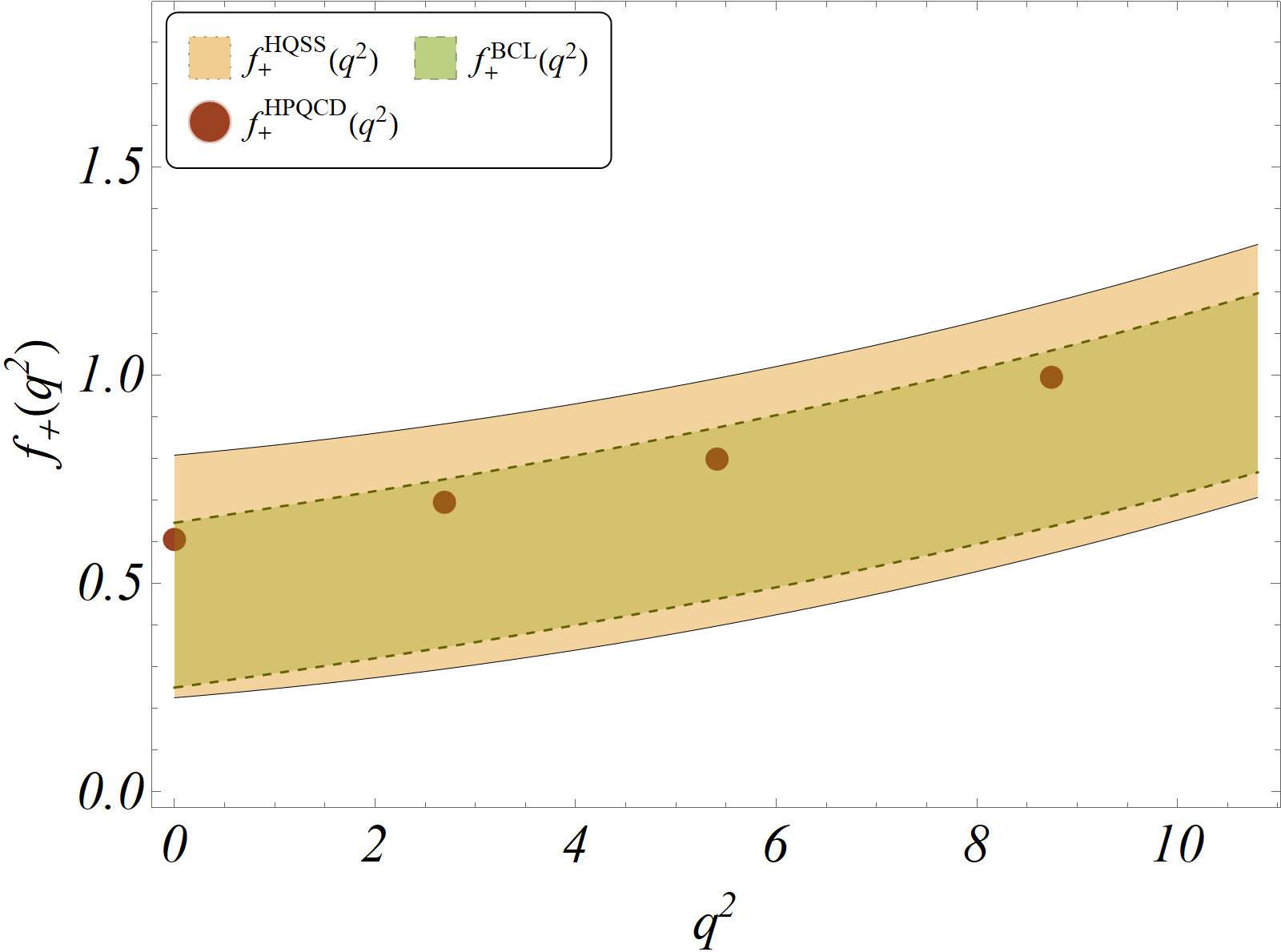}\label{fig:fplufig}}~~~~~~
	\subfloat[]{\includegraphics[scale=0.35]{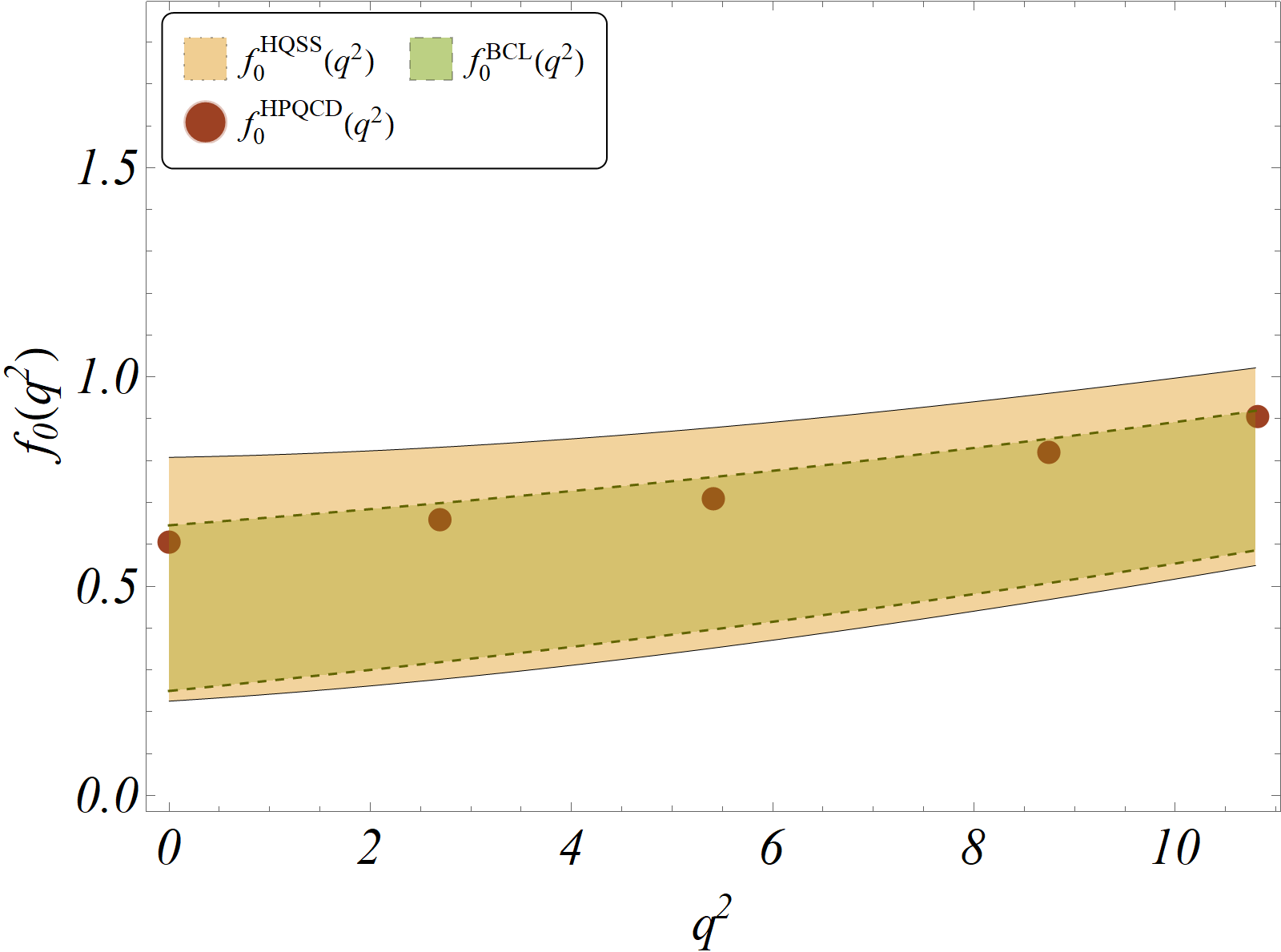}\label{fig:f0fig}}
	\caption{The shapes for the $B_c\rightarrow \eta_{c}$ form factors computed with the fit results displayed in table~\ref{tab:HQSSfitparam} (third column) and \ref{tab:BCLcoeffhqss}, respectively. We overlay the data provided by the HPQCD collaboration\cite{Colquhoun:2016osw}. The bands represent the $1\sigma$ region for the corresponding form factors. The brown band representing the form factors corresponding to the HQSS parametrization and the green band specifying the BCL parametrization using the fit results from table~\ref{tab:BCLcoeffhqss}.}
	\label{fig:Bc2etacplot}
\end{figure*}

Our objective is to obtain the $q^2$ shapes of the form factors $f_{+,0}^{B_c\to \eta_c}(q^2)$, which in the HQSS framework is parametrized as shown in the relations provided in eqs.~\ref{eqn:FFrelns} and~\ref{eqn:Delta}, respectively. We need the inputs on $\Delta(1)'$, $\Delta(1)''$, $\Delta_{+}(1)$ and $\Delta_0(1)$ respectively in order to obtain these shapes. The inputs on the parameters $\Delta(1)'$ and $\Delta(1)''$ can be collected from the fit results provided in the third column of table~\ref{tab:HQSSfitparam}. In order to fix the values of $\Delta_{+}(1) = \Delta_0(1) = \Delta$ we use the HQSS relation given in eq. \ref{eq:hqssw1}. To obtain $\Delta$, we use an average over the values of $\Delta_V(1)$, $\Delta_{A_1}(1)$ and $\Delta_{A_0}(1)$ obtained from the fit (the results of which we have shown in the third column of table~\ref{tab:HQSSfitparam}). As a symmetry-breaking correction, we have added an additional error of $30\%$ to the estimate of $\Delta$. Following this method we obtain
\begin{equation}\label{eq:deltap0n}
\Delta_{+}(w=1) = \Delta_0(w=1)= \Delta =  0.84 \pm 0.04 \pm 0.25.
\end{equation}
Here, the first contribution to the estimated error comes from the inputs used in the fit, and the second contribution is added as the breaking of the HQSS symmetry. The corresponding shapes of $f_{+}(q^2)$ and $f_0(q^2)$ are shown in figs.~\ref{fig:fplufig} and \ref{fig:f0fig} respectively. We also overlay the data points from ref.~\cite{Colquhoun:2016osw} over the bands we obtain for $f_{0,+}^{B_c\to\eta_c}$. However, without any information regarding the errors on the said data points, an objective statistical comparison between our results and those from ref.~\cite{Colquhoun:2016osw} is currently unwarranted.

To get the shape of the form factors in HQSS, the parametrizations for the $\Delta_i(w)$ are given in eq. \ref{eqn:Delta}, which is based on a Taylor series expansion about the zero recoil $w=1$ or the maximum value of $q^2$. This suggests that this expansion will provide a more reliable result near $w=1$. Therefore, to get an appropriate estimate of the $q^2$ shapes of the form factors, we have adopted a model-independent approach proposed by Bourrely-Lellouch-Caprini (BCL) to parametrize the $q^2$ shapes of $f_{+,0}(q^2)$ \cite{Caprini, Leljak:2019eyw}. In this parametrization, both the form factors are written as a polynomial series in powers of $q^2$.
\begin{equation}
f_i(q^2) = \frac{1}{P(q^2)} \sum_{k=0}^{n} a_i^k z^k( q^2,t_0)
\label{eqn:BCLeta}
\end{equation}
with $i=0,+$ and
\begin{eqnarray}
z(q^2,t_0) &=& \frac{ \sqrt{t_+ - q^2} - \sqrt{t_+ - t_0}}{\sqrt{t_+ - q^2} + \sqrt{t_+ - t_0}}, \nonumber\\
t_0 &=&  t_+  \left (1 - \sqrt{1 - \frac{t_-}{t_+}} \right ).
\label{eqn:zexp}
\end{eqnarray}
\begin{table*}[t]
	\begin{center}
		\begin{tabular}{|*{6}{c|}}
			\hline
			\multicolumn{6}{|c|}{Pseudo data points for $f^{B_c\to \eta_c}_{+,0}(q^2)$} \\
			\hline
			$f_{+}(6)$ &  	$f_{+}(8)$ & $f_{+}(\text{$q^2_{max}$})$ & $f_{0}(6)$ & $f_{0}(8)$ & $f_{0}(\text{$q^2_{max}$})$ \\
			$\text{0.72(28)}$& $\text{0.83(30)}$& $\text{1.01(30)}$& $\text{0.63(26)}$ & $\text{0.69(25)}$ &  $\text{0.79(24)}$ \\
			\hline
		\end{tabular}
		\caption{Synthetic data for the $B_c\rightarrow \eta_c$ form factors generated using eq.~\ref{eqn:Delta} and the results of $\Delta(1)'$ and $\Delta(1)''$ given in the third column of tab~\ref{tab:HQSSfitparam}. Also, to generate the numbers, we have used the conservative inputs $\Delta_{+}(w=1)= \Delta_0(w=1)= 0.84(25)$ in eq.~\ref{eq:deltap0n}. We have presented the respective correlations in table \ref{tab:corrfp0}.}
		\label{tab:syndatfpf0}
	\end{center}
\end{table*}
\begin{table*}[t]
	\begin{center}
		\begin{tabular}{|*{3}{c|}}
			\hline
			\text{BCL coefficients} & \text{Fit Results}   \\
			\hline
			$a_0^0$  &  $\text{0.51(16)}$\\
			$a_0^1$  &  $\text{-3.7(22)}$  \\
			$a_+^0$  &  $\text{0.58(18)}$\\
			$a_+^1$  &  $\text{-8.0(13)}$ \\
			\hline
			$\text{dof}$  &  $2$ \\
			$\text{p-Value}$  & $0.13$ \\
			\hline
		\end{tabular}
		\caption{Fit results for the BCL coefficients $a_i^n$ corresponding to the $B_c\rightarrow \eta_{c}$ transition form factors obtained from a fit to the synthetic data generated under HQSS displayed in table~\ref{tab:syndatfpf0}. We truncate the series at n=2 for $f_0$ and n=1 for $f_+$, since higher order coefficients are insensitive to the data. Correlation between the parameters presented in table \ref{tab:corrBCLcoeff}.}
		\label{tab:BCLcoeffhqss}
	\end{center}
\end{table*}

In the above, $t_{\pm}=(m_{B_c}\pm m_{\eta_c})^2$ and $P(q^2) = 1- q^2/m_R^2$. $m_R$ represents the masses of the low-lying $B_c$ resonances. We use $m_R = 6.331(7)$ GeV and $6.712(19)$ GeV for $f_+$ and $f_0$ respectively in accordance with ref.~\cite{Mathur:2018epb}. The goal here is the extraction of the BCL expansion coefficients $a_i$'s. To that end, we first generate pseudo data points for $f_{+,0}(q^2)$ following eqs.~\ref{eqn:FFrelns} and \ref{eqn:Delta} respectively using the fit results of $\Delta(1)'$ and $\Delta(1)''$ given in third column of table~\ref{tab:HQSSfitparam} and $\Delta_{+}(1)$, $\Delta_0(1)$ from eq.~\ref{eq:deltap0n} and their correlation given in table \ref{tab:corrHQSSparm2}.

These data points are provided in table \ref{tab:syndatfpf0}, and can be compared to more precise lattice estimates in the future. Note that to extract the BCL coefficients, we have generated pseudo data points at $q^2 = 6, 8, \text{$q^2_{max}$}$ GeV$^2$. These points are very close to the maximum allowed value of $q^2$ where the HQSS expansion is relatively reliable as compared to the low $q^2$ regions. We have checked that we get identical results using the results in the second column of table~\ref{tab:HQSSfitparam}, which we have not shown. We then fit the BCL coefficients using these pseudo data points, the results of which we have presented in table~\ref{tab:BCLcoeffhqss}. We reduce the number of free parameters from five to four by using the QCD relation $f_+(0)=f_0(0)$. Note that in the physically allowed region for $q^2$, $|z(q^2,t_0)|$ is $\lsim 0.02$, enabling us to constrain the $q^2$ shapes of $f_{+,0}(q^2)$ with a very minimal number of coefficients ($a^k_i$). Higher-order coefficients are multiplied by the higher power of the $z$ values ($z^k<< 1$). Hence overall contributions from the higher-order terms in the series will be negligibly small compared to the leading-order terms. In our analysis, we have found that the coefficients for $k \ge 2$ are not sensitive to our inputs and have no impact on our final results. The respective $q^2$ shapes obtained using these fit results and the related correlations are shown in figs.~\ref{fig:fplufig} and \ref{fig:f0fig} respectively. As expected, these agree with the one obtained using HQSS expansion. We obtain the following values at $q^2=0$:
\begin{equation}\label{eq:fbcetacq20}
f_{+}^{B_c\to \eta_c}(q^2=0) = f_{0}^{B_c\to \eta_c}(q^2=0) =
\begin{cases}
0.52 \pm 0.28 \ \ \ \text{(HQSS)},\\
 0.45 \pm 0.20 \ \ \ \text{(BCL)}.
\end{cases}
\end{equation}
Note that the estimated values for $f_+(0)(=f_0(0))$ using HQSS and BCL parametrization are in agreement within their respective uncertainties. However, the estimated error in HQSS is about 53\% while that in the BCL is about 44\%. Following the discussion above, to predict the observables related to $B_c^-\to \eta_c\ell^-\bar{\nu}$ we will use the $q^2$ shapes of the form factors obtained using the BCL parametrization.

\begin{figure}[t!]
	\centering
	{\includegraphics[scale=0.42]{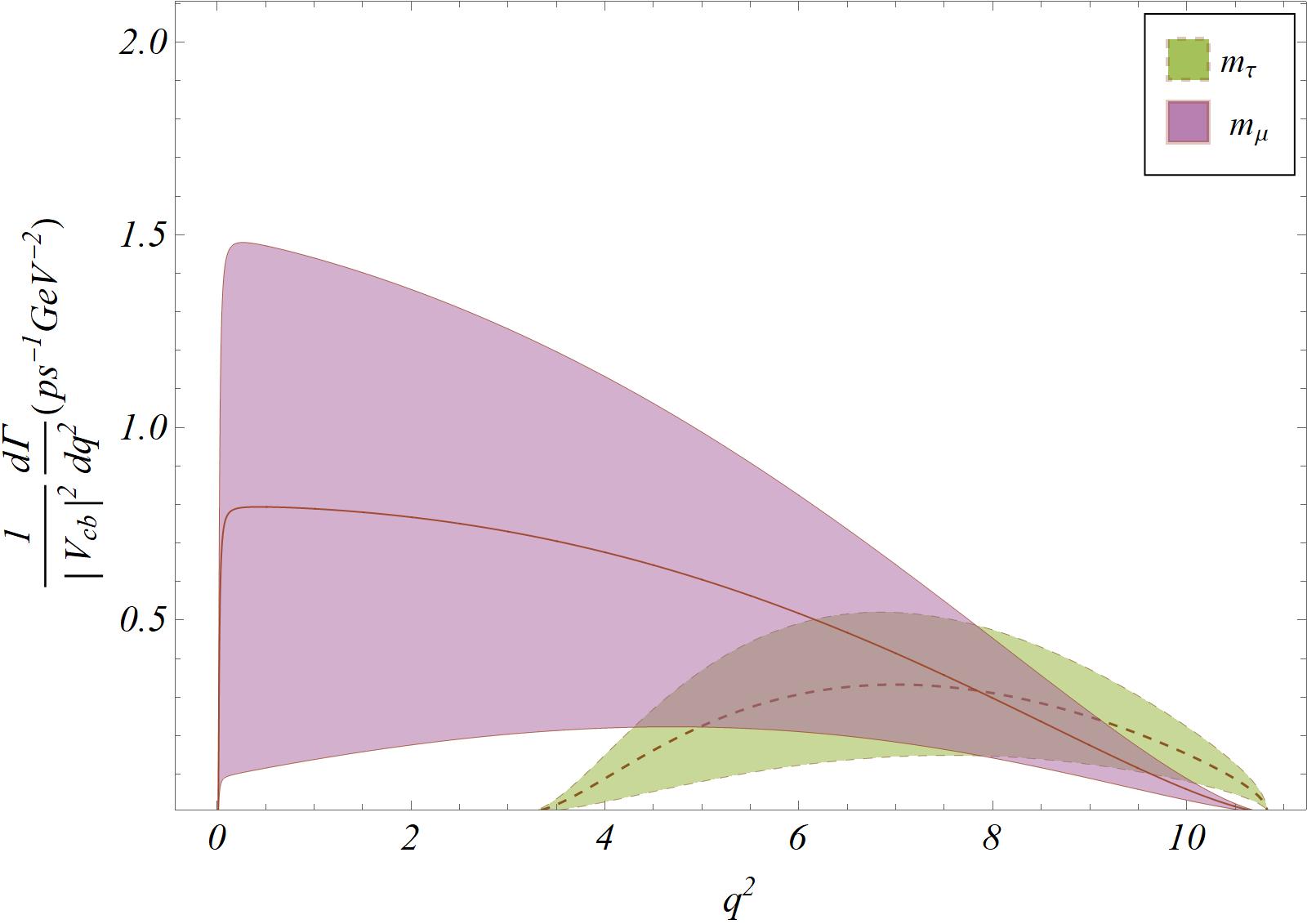}\label{fig:drates}}~~~~~~
	\caption{1$\sigma$ bands for the differential decay distributions corresponding to semileptonic $B_c\to\eta_c$ decays with a tauon (green) or a muon (purple) in the final state.}
	\label{fig:bc2etacdrates}
\end{figure}

Using the estimates on the shape of $f_{+,0}(q^2)$ we have predicted the shape of $\frac{1}{|V_{cb}|^2} \frac{d\Gamma(B_c^-\to \eta_c\ell^-\bar{\nu})}{ dq^2}$ with $\ell = \tau, \mu(e)$, and have shown them in fig.~\ref{fig:bc2etacdrates}. The overall normalization is dependent on the value of $|V_{cb}|^2$. Our estimates for the uncertainties are rather large at present. We expect this to improve with more precise inputs on the form factors so that the shape can be compared to future experiments.
We also provide the $q^2$ integrated branching fraction $\mathcal{B}(B_c^-\to \eta_c\ell^-\bar{\nu})$:
\begin{equation}
\frac{1}{|V_{cb}|^2} \mathcal{B}(B_c^-\to \eta_c\tau \bar{\nu}) =  (0.94 \pm 0.49), \ \ \ \  \frac{1}{|V_{cb}|^2} \mathcal{B}(B_c^-\to \eta_c\mu^- \bar{\nu}) = (3.13 \pm 1.97).~~~
\end{equation}
With $|V_{cb}| = 0.0403(5)$ \cite{Ray:2023xjn} we obtain for the SM
\begin{equation}
\mathcal{B}(B_c^-\to \eta_c\tau^- \bar{\nu}) =  (1.53 \pm 0.80)\times 10^{-3}, \ \ \ \   \mathcal{B}(B_c^-\to \eta_c\mu^- \bar{\nu}) = ( 5.08 \pm 3.21)\times 10^{-3}.~~~~
\end{equation}
The predicted values of the semileptonic branching fractions have an error of around 52\%. Here, the major sources of errors are the form factors. $|V_{cb}|$ has an error about 1\%.

Also, using this result for the form factors, we have predicted the lepton flavor universality conserving observable $R(\eta_c)=\frac{\Gamma(B_c^-\rightarrow \eta_c \tau\bar{\nu}_{\tau})}{\Gamma(B_c^-\rightarrow \eta_c \mu^-\bar{\nu}_{\mu})}$ which is given by
	\begin{align}
	R(\eta_c)=0.310(42).
	\end{align}
The estimated error in the ratio is about $13\%$ though the form factors accross the $q^2$ region have an error roughly about 40\%. The reduction of error in $R(\eta_c)$ is due to a high positive correlation between the form factors in the numerator and the denominator, which we have explicitly checked\footnote{The standard deviation of a ratio $R = A/B$ of two observables A and B is given by
\begin{equation}
\sigma_{R} = |R| \sqrt{\left(\frac{\sigma_{A}}{A}\right)^2 + \left(\frac{\sigma_{B}}{B}\right)^2 - 2 \rho_{AB} \ \frac{\sigma_{A}}{A} \frac{\sigma_{B}}{B} }
\end{equation}
where $\rho_{AB}$ is the correlations between the uncertainties of A and B. Here, $\sigma_A$ and $\sigma_B$ represent the standard deviations in the estimate of $A$ and $B$ respectively.}. Our prediction can be compared to the earlier predictions from other references in the literature:
\begin{equation}
R(\eta_c) =
\begin{cases}
  0.31 ^{+0.04}_{- 0.02}\ \ \ \  \text{\cite{Murphy:2018sqg}},\\
 0.30 \pm 0.05\ \  \text{\cite{Cohen:2019zev}}, \\
 0.29 \pm 0.05\ \ \text{ \cite{Berns:2018vpl}}.
\end{cases}
\end{equation}
Note that our prediction agrees with all these predictions. All though these three predictions have errors comparable to ours but we do carry out a more comprehensive study using the available lattice inputs. None of these analyses use lattice input on $B_c\to J/\psi$ form factors. They have used the HPQCD \cite{Colquhoun:2016osw} preliminary lattice points on $f_{+,0}^{B_c\to \eta_c}$ as inputs in their analyses and added an error of about 20\% without any correlations among them. On the other hand, our lattice pseudo data points for $f_{+,0}^{B_c\to \eta_c}$ are correlated, because of which we will get a reasonably strong correlation between the BCL coefficients of the of $f_+(q^2)$ and $f_0(q^2)$. This will lead to a strong correlation between the numerator and the denominator of $R(\eta_c)$ ratio, which essentially reduces the error in $R(\eta_c)$. Our results could be improved with the available higher order correction to the HQSS relations. All these predictions can be tested in future experiments, which will be helpful in understanding the predictive power of HQSS.

\section{Extractions of $\psi^{R}_{M}(0)$ for $M = B_c, J/\psi, \eta_c$
}\label{sec:S Wave}

As we have mentioned in the introduction, a quantitative knowledge of the charmonium ($\psi$ and $\eta_c$) and $B_c$ meson radial wave functions at the origin is imperative in order to obtain estimates for exclusive dynamics (decays, productions, etc.) involving such mesons. In this section, we will discuss the analysis of the extractions. Apart from the lattice, to date, several theoretical approaches have been incorporated in order to estimate the $B_c\to J/\psi (\eta_c)$ form factors such as perturbative QCD (PQCD)~\cite{Wang:2012lrc, PhysRevD.90.114030, Sun:2008ew,Rui:2016opu, Liu:2023kxr}, the
constituent quark model~\cite{Anisimov:1998uk}, the relativistic quark model~\cite{Nobes:2000pm, Ebert:2003cn, Ivanov:2005fd, Ebert:2010zu, Nayak:2022gdo}, the non-relativistic quark model~\cite{Hernandez:2006gt, Li:2019tbn}, QCD sum rules~\cite{Colangelo:1992cx, Kiselev:2002vz, Leljak:2019eyw, Azizi:2009ny, Azizi:2013zta, Kiselev:1999sc}, the relativistic constituent quark model~\cite{Ivanov:2006ni}, the light-front covariant quark model (LFCQ)~\cite{Wang:2008xt, Ke:2013yka}. In many of these model analyses, the form factors calculated are sensitive to $\psi(0)_{B_c}$ and $\psi(0)_{J/\psi(\eta_c)}$. We attempt to estimate the non-perturbative matrix elements without assuming a dynamical model, in a data driven way.

In this analysis, we will use the results of the calculation of the form factors in the framework of NRQCD, which is an effective theory (EFT) approach \cite{Brambilla:2004jw, PhysRevD.51.1125, Bodwin:2006dm, Brambilla:2010cs, Kiselev:2001zb, Bell:2005gw, Bell:2006tz, Qiao:2011yz, Qiao:2012vt, Qiao:2012hp, Zhu:2017lqu, Zhu:2017lwi, Tao:2022yur}.  For large enough masses, the heavy quark-antiquark system can be treated as non-relativistic (NR). Such a system involves different scales: hard (mass of the heavy quark $m$), soft (the relative momentum $m v$ of the quark-antiquark pair) and the ultra-soft (the binding energy $E =m v^2$) with $v << 1$ and the hierarchy: $m >> m v >> m v^2$. Another scale which is relevant for the discussion is the QCD scale $\Lambda_{QCD}$. An important fact regarding NRQCD that separates it from other approaches is that it can be systematically derived from the QCD Lagrangian. NRQCD is obtained by integrating out the hard scale $m$ from the QCD Lagrangian. The only assumption required is $m >> \Lambda_{QCD}$. The important non-perturbative physics involves momenta of order $m v$ and less. The relativistic effects are separated from the non-relativistic effects. However, the relativistic states (light degrees of freedom) impact the low energy physics. The relevant interactions are local and systematically added as corrections to the leading order term of the NRQCD Lagrangian as a power series in $1/m$. As a result, the effective Lagrangian can be expressed as a series expansion in $\alpha_s$ and $1/m$:
\begin{equation}
\mathcal{L}_{NRQCD}= \sum_{n} \frac{C_n(\alpha_s(m),\mu)}{m^n} \mathcal{O}_n(m v, m v^2, \mu).
\end{equation}
Here, $\mathcal{O}_n$'s are the low-energy effective operators and $C_n$'s are the corresponding perturbatively calculable Wilson coefficients (WCs). $\mu$ is the factorization scale.

\begin{figure}[htb!]
	\centering
	\begin{tikzpicture}[baseline={(current bounding box.center)}]
	\begin{feynman}
	\vertex(a);
	\node[right=2cm of a, crossed dot, style=black](a1);
	\vertex[right=4cm of a] (a2);
	\vertex[left=0.6cm of a2] (a3);
	\vertex[below=2cm of a] (b);
	\vertex[right=4cm of b] (c1);
	\vertex[left=0.6cm of c1] (c2);
	\diagram* {
		(a) -- [anti fermion ,arrow size=1pt,edge label={\(\overline{b}\)}] (a1) -- [anti fermion ,arrow size=1pt,edge label={\(\overline{c}\)}] (a2) ,(b) --[fermion ,arrow size=1pt,edge label'={\(c\)}] (c1),(a)--[fill=gray!15,plain,bend left](b),(a)--[fill=gray!15,plain,bend right,edge label'={\(B_c^+\)}](b),(a2)--[fill=gray!15,plain,bend left,edge label={\(J/\psi(\eta_c)\)}](c1),(a2)--[fill=gray!15,plain,bend right](c1),(a3)--[gluon](c2)
	};
	\end{feynman}
	\end{tikzpicture}
	\begin{tikzpicture}[baseline={(current bounding box.center)}]
	\begin{feynman}
	\vertex(a);
	\node[right=2cm of a, crossed dot, style=black](a1);
	\vertex[right=4cm of a] (a2);
	\vertex[right=0.75cm of a] (a3);
	\vertex[below=2cm of a] (b);
	\vertex[right=4cm of b] (c1);
	\vertex[right=0.75cm of b] (c2);
	\diagram* {
		(a) -- [anti fermion ,arrow size=1pt,edge label={\(\overline{b}\)}] (a1) -- [anti fermion ,arrow size=1pt,edge label={\(\overline{c}\)}] (a2) ,(b) --[fermion ,arrow size=1pt,edge label'={\(c\)}] (c1),(a)--[fill=gray!15,plain,bend left](b),(a)--[fill=gray!15,plain,bend right,edge label'={\(B_c^+\)}](b),(a2)--[fill=gray!15,plain,bend left,edge label={\(J/\psi(\eta_c)\)}](c1),(a2)--[fill=gray!15,plain,bend right](c1),(a3)--[gluon](c2)
	};
	\end{feynman}
	\end{tikzpicture}
	\caption{The leading order Feynman diagram for semileptonic decays $B_c^+\to J/\psi(\eta_c)\ell^+\nu_{\ell}$}
	\label{fig:feynman}
\end{figure}
In full-QCD, the $B_c\rightarrow J/\psi(\eta_{c})$ transition form factors $f_+$, $f_0$, V, $A_0$, $A_1$, $A_2$ are defined as follows:
\begin{align}\label{eq:qcdffdef}
\langle \eta_{c}(p)|\bar{c}\gamma^{\mu}b|B_c(P)\rangle =&f_{+}(q^2)\bigg[P^{\mu}+p^{\mu}-\frac{m_{B_c}^2-m_{\eta_{c}}^2}{q^2} q^{\mu}\bigg]+f_{0}(q^2)\frac{m_{B_c}^2-m_{\eta_{c}}^2}{q^2} q^{\mu},\nonumber\\
\langle J/\psi(p,\epsilon^*)|\bar{c}\gamma^{\mu}b|B_c(P)\rangle =&\frac{2 i V(q^2)}{m_{B_c}+m_{J/\psi}} \epsilon^{\mu \nu \rho\sigma}\epsilon^*_{\nu}p_{\rho}P_{\sigma},\nonumber\\
\langle J/\psi(p,\epsilon^*)|\bar{c}\gamma^{\mu}\gamma_{5}b|B_c(P)\rangle =&2 m_{J/\psi}A_{0}(q^2)\frac{\epsilon^* \cdot q}{q^2}q^{\mu}-A_2(q^2)\frac{\epsilon^*\cdot q}{(m_{B_c}+m_{J/\psi})}\bigg[P^{\mu}+p^{\mu}-\frac{m_{B_c}^2-m_{\JP}^2}{q^2} q^{\mu}\bigg]\nonumber\\&+ (m_{B_c}+m_{J/\psi})A_1(q^2)(\epsilon^{*\mu}-\frac{\epsilon^*\cdot q}{q^2}q^{\mu}).
\end{align}
Where the momentum transfer $q=P-p$.
The $B_c$ meson is a bound-state consisting of two heavy quarks with different flavors; the masses of which are larger than $\Lambda_{QCD}$. The relative velocity of these heavy quarks within the $B_c$ meson is small, i.e. $v<<1$. However, the magnitude is still considerably more than the velocities of constituent quarks in the final state charmonium. One can hence apply the non-relativistic QCD formalism in order to study the  semileptonic $B_c$ decays to charmonium. In the NRQCD formalism, the matrix element relevant to the $B_c\to J/\psi (\eta_c)$ form-factors can be factorised as
\begin{eqnarray}\label{eq:NRQCDBCtoJpsiff}
\langle J/\psi(\eta_{c}) |\bar{c}\gamma^{\mu}(1-\gamma_{5})\nu| B_c\rangle\simeq\sum_{n=0} \psi(0)_{B_c}\psi(0)_{J/\psi(\eta_c)} T^n.
\end{eqnarray}
Here, the nonperturbtive parameters $\psi(0)_{B_c}$, $\psi(0)_{J/\psi(\eta_c)}$ are the Schr{\"o}dinger wave functions at the origin for the b$\bar{c}$ and c$\bar{c}$ systems, which are defined as
\begin{align}
\psi(0)_{\eta_c}  &= \frac{1}{\sqrt{2 N_c}} \langle\eta_{c}|\psi_{c}^{\dagger}\chi_c|0\rangle, \nonumber \\
\psi(0)_{B_c} & = \frac{1}{\sqrt{2 N_c}} \langle0|\chi_b^{\dagger}\psi_{c}|B_c\rangle, \nonumber \\
\psi(0)_{J/\psi} &= \frac{1}{\sqrt{2 N_c}} \langle J/\psi|\psi^{\dagger}\chi_c|0\rangle.
\label{eq:nonpert_matrix}
\end{align}
The hard scattering kernels $T^n$ can be calculated perturbatively. Here, both the $B_c$ meson and the charmonium are treated as non-relativistic bound states. The leading order ($\mathcal{O}(\alpha_s)$) results are obtained from the diagrams in Fig.~\ref{fig:feynman} \cite{Bell:2006tz}. The next-to-leading order (NLO) corrections at $\mathcal{O}(\alpha_s^2)$ accuracy are also available in the literature \cite{Bell:2006tz,Qiao:2012hp,Qiao:2012vt}. The relativistic corrections to these form-factors, which are calculated in ref.~\cite{Zhu:2017lqu} play an essential role in the phenomenology of $B_c$ meson decays to charmonium. These corrections are separated into two parts since there are two bound states composed of a heavy quark and a heavy antiquark: the charmonium and the $B_c$ meson. We define $v$ as the relative velocity between the quarks inside the quarkonium and $v'$ as that inside the $B_c$ meson. Half of the relative momentum the quark inside the charmonium is defined as $k=\frac{m_c v}{2}$ and half of the quark relative momentum is defined as $k'=m_{red} v'=m_b m_c v'/(m_b+m_c)$ inside the $B_c$ meson. Accordingly, the masses of the charmonium bound states  and $B_c$ will be defined.
In ref. \cite{Zhu:2017lqu}, the authors expand the amplitude in powers of $k^{\mu}$ in order to calculate the relativistic corrections from the charm quark-antiquark pair interactions inside the charmonium. Analogously, the relativistic corrections to the form factors from the charm and bottom quark interactions inside the $B_c$ meson can be obtained when the amplitude is expanded in powers of $k^{\prime \mu}$. To estimate the magnitude of the relativistic correction operator matrix elements, one has
\begin{align}\label{eq:matrixelements2}
\langle\eta_{c}|\psi_{c}^{\dagger}\bigg(-\frac{i}{2}\mathcal{D}\bigg)^2\chi_c|0 \rangle &\simeq |\boldmath k|^2 \langle \eta_{c}|\psi_{c}^{\dagger}\chi_c|0\rangle = |k|^2 \psi(0)_{\eta_c}\nonumber \\
\langle 0|\chi_b^{\dagger}\bigg(-\frac{i}{2}\mathcal{D}\bigg)^2\psi_c|B_c\rangle &\simeq |\boldmath k'|^2 \langle 0|\chi_b^{\dagger}\psi_{c}|B_c\rangle =  |\boldmath k'|^2 \psi(0)_{B_c} \nonumber \\
\langle J/\psi|\psi_{c}^{\dagger}\bigg(-\frac{i}{2}\mathcal{D}\bigg)^2\chi_c|0 \rangle &\simeq |\boldmath k|^2 \langle J/\psi|\psi^{\dagger}\chi_c|0\rangle = |\boldmath k|^2 \psi(0)_{J/\psi}
\end{align}
where,
\begin{center}
	$|v|_{J/\psi(\eta_c)}^2$$ \approx 0.201$, $~~$ $|v|_{B_c}^2\approx0.186$.\\
\end{center}
Hence, the matrix elements of the relativistic operators can be estimated by the wave functions at the origin of the heavy quarkonium. In ref.~ \cite{Zhu:2017lqu}, including all these corrections as mentioned above the transition form-factors are expressed as
\begin{eqnarray}
F_i(q^2)=F_i^{LO}(q^2)\bigg[1+\frac{\alpha_{s}}{4\pi} f_{i,\alpha_s}(s,\gamma) +\frac{m_c^2 |v|^2 }{4 m_b^2} f_{i,RC}(z,y) + \frac{m_c^2 |v^{\prime}|^2 }{(m_b+m_c)^2}  f_{i, {RC}^{\prime}}(z,y) + h.o \bigg],
\end{eqnarray}
where, $s=m_b^2/(m_b^2-q^2)$, $\gamma = (m_b^2-q^2)/(4m_b m_c)$, $z = m_c^2/m_b^2$ and $y = \sqrt{q^2/m_b^2}$ respectively. The various functions $F_i^{LO}(q^2)$, $f_{i,\alpha_s}(s,\gamma)$, $f_{i,RC}(z,y)$, and $f_{i, {RC}^{\prime}}(z,y)$ are obtained from the refs. \cite{Qiao:2012hp,Zhu:2017lqu}. The leading order contributions $F_i^{LO}(q^2)$ are proportional to the product of $\psi(0)_{B_c} \psi(0)_{J/\psi(\eta_c)}$. Note that in the limit $m_b \to \infty$, the relativistic corrections can be expressed in leading power in $z=\frac{m_c}{m_b}$ and the only unknown parameters are the non-perturbative matrix elements as defined in Eq.~\ref{eq:nonpert_matrix} which need to be extracted from the available inputs. As mentioned in the introduction, we fit them using the available lattice inputs \cite{Harrison:2020nrv} on the $B_c\to J/\psi$ form-factors and a few other inputs. To obtain the form factors in NRQCD, we use the inputs on the strong coupling constant $\alpha_s$ and the quark masses $m_b$ and $m_c$ following the kinetic, $\overline{MS}$, and pole-mass scheme. The corresponding values are given in table~\ref{tab:massscheme}. With these inputs, we have estimated the relative size of the perturbative and relativistic corrections in all schemes. We found that the corrections are large and highly scheme dependent. The details will be discussed in the following section.
Note that the radial part of the Schr{\"o}dinger wave functions defined in Eq.~\ref{eq:nonpert_matrix} are expressed as $\psi^R_{M}(0)$ via the relation

\begin{equation}
\psi(0)_M = \frac{1}{\sqrt{4\pi}}\psi_M^R(0).
\label{eq:schradial}
\end{equation}

The factor of $\frac{1}{\sqrt{4\pi}}$ comes from the spherical harmonics $Y_l^m(\theta,\phi)$ as a result of assuming a spherically symmetric potential.
The radial part $\psi_M^R(0)$ arises as solutions to two-body non-relativistic quantum-systems at the origin (i.e $r=0$). In this paper, we will present our fit results for the radial wave functions: $\psi_{B_c}^R(0)$, $\psi_{J/\psi}^R(0)$ and $\psi_{\eta_c}^R(0)$, respectively. The corresponding Schr{\"o}dinger wave functions can be obtained using the relation given in Eq.~\ref{eq:schradial}.

Note that apart from the form factors we have discussed above the decay constants $f_{J/\psi}$, $f_{\eta_c}$ and $f_{B_c}$ are also sensitive to the radial wave functions
$\psi_{J/\psi}^R(0)$, $\psi_{\eta_c}^R(0)$ and $\psi_{B_c}^R(0)$, respectively.
The analytical expressions of these decay constants calculated in the NRQCD framework are given by
	\begin{eqnarray}\label{eq:fpsi}
	f_{J/\psi}= \sqrt{\frac{3}{ 2 m_c \pi}} \psi^R_{J/\psi}(0) \left[1 - \frac{8}{3} \, \frac{\alpha_s}{ \pi}-\frac{1}{6} \langle v^2 \rangle_{J/\psi}+\frac{\alpha_s}{3 \pi}\bigg(\frac{8}{9}+\frac{8}{3} \ln\frac{\mu_{\Lambda}^2}{m_c^2}\bigg)\langle v^2 \rangle_{J/\psi}+\frac{29}{18}\langle v^4 \rangle_{J/\psi}\right],
	\end{eqnarray}
	\begin{align}\label{eq:fetac}
	f_{\eta_{c}} = \sqrt{\frac{3}{ 2 m_c \pi}} \psi^R_{\eta_{c}}(0)& \Bigg[ 1+\frac{\alpha_s}{\pi} \frac{\pi^2-20}{3} + \langle v^2 \rangle_{\eta_{c}} \left(-\frac{4}{3}+
	\frac{\alpha_s}{\pi}\frac{1}{27}(48\ln\frac{\mu_{\Lambda}^2}{m_c^2}-96\ln 2-15\pi^2+196)\right) \\ \nn
	& +\frac{68}{45}\langle v^4 \rangle_{\eta_{c}}\Bigg]^{1/2}
	\end{align}
	and
	\begin{equation}\label{eq:fBc}
f_{B_c}=\sqrt{\frac{2}{m_{B_c}}}\Bigg[c_0^f \langle 0|\chi_c^{\dagger}\psi_{b}|\overline{B_c}(P)\rangle +\frac{c_2^f}{m_{B_c}^2} \langle 0|\chi_c^{\dagger}\bigg(-\frac{i}{2}\mathcal{D}\bigg)^2 \psi_{b}|\overline{B_c}(P)\rangle +\mathcal{O}(v^4) \Bigg].
\end{equation}
The QCD corrections at order $\alpha_s$ and the relativistic corrections at order $v^2$, $v^4$ and $\alpha_s v^2$ are known. In the above expressions, we choose the factorization scale $\mu_{\Lambda}=m_c$.

We have taken the analytical expressions for $f_{J/\psi}$ from the ref.~\cite{Bodwin:2002cfe, Lee:2018aoz} and that for $f_{\eta_c}$ from \cite{Guo:2011tz, Bodwin:2002cfe}. The expression for the decay constant $f_{B_c}$ have been taken from the ref.~\cite{Wang:2015bka, Bodwin:2002cfe}. In eq.~\ref{eq:fBc}, the coefficients $c_0^{f}$, $c_2^{f}$ are taken from ref~\cite{Wang:2015bka} and the matrix elements are defined in eqs.~\ref{eq:nonpert_matrix} and \ref{eq:matrixelements2} respectively. It is evident from these equations that the decay constant defined in eqs.~\ref{eq:fBc} is sensitive to the mesonic wave function $\psi^R_{B_c}(0)$.

\subsection{Available inputs}\label{subsec:inputs}
As discussed in the last section, the transition form factors in NRQCD are parametrized in terms of the mesonic wave functions: $\psi_{B_c}^R(0)$, $\psi_{J/\psi}^R(0)$ and $\psi_{\eta_c}^R(0)$. Furthermore, the decay constants which we have described above are also sensitive to the same radial wave functions. We will extract them using the available lattice inputs and the experimental data. In what follows, we specify the theory as well as experimental inputs that we use in our analysis to extract the radial wave functions.

Note that the form-factor estimates in NRQCD are more reliable near $q^2=0$. In table \ref{tab:inputlattQCDSR}, we have shown the available lattice inputs on $B_c\to J/\psi$ form factors. Furthermore, in section \ref{sec:Bctoetac} we have extracted the information on the $B_c \to \eta_c$ form factors. In the same table, we have presented the lattice estimates on the decay constants $f_{J/\psi}$, $f_{\eta_c}$ and $f_{B_c}$.

	\begin{table*}[t!]
		\begin{center}
  \resizebox{0.7\textwidth}{!}{
			\begin{tabular}{|*{4}{c|}}
				\hline
				\text{\bf Form factors} $(q^2)$& \text{\bf Lattice (Syn.)}~\cite{Harrison:2020gvo} & \text{\bf Decay constant} & \textbf{Lattice estimates}\\
				&&& (in MeV)\\
				\hline
				$V^{J/\psi}(0)$ & $0.725(68)$ &	$f_{B_c}$  & 434 (15)~\cite{McNeile:2012qf, Colquhoun:2015oha}   \\
		$A_{0}^{J/\psi}(0)$ & $0.477(32)$ & $f_{J/\psi}$ & 405(6)~\cite{Donald:2012ga}  \\
		$A_{1}^{J/\psi}(0)$ & $0.457(29)$ & $f_{\eta_{c}}$ & 394.7(2.4)~\cite{Davies:2010ip}  \\
		$A_{2}^{J/\psi}(0)$ & $0.417(88)$ & & \\
				\hline
			\end{tabular}
   }
			\caption{Values for the $B_c\to J/\psi$ form factors at the origin estimated from Lattice QCD~\cite{Harrison:2020gvo} and the lattice estimates on the decay constants for $J/\psi$, $\eta_c$ and $B_c$ mesons.}
			\label{tab:inputlattQCDSR}
		\end{center}
	\end{table*}

On the experimental side, we use the Branching Ratios (BR's) of the decays: $\eta_{c}\rightarrow \gamma\gamma$ and $J/\psi\rightarrow e^+ e^- $ in our analysis as input in order to constrain the parameters. In particular these inputs are sensitive to $\psi^R_{J/\psi}(0)$ and $\psi^R_{\eta_c}(0)$ respectively. Hence, the data will be helpful to constrain the relevant parameters which in turn will help constrain $\psi^R_{B_c}(0)$. Details about the theoretical expressions for these decays are as follows:
\begin{itemize}
	\item \underline{$\eta_{c}\rightarrow \gamma\gamma$}:\\
	The analytic expression for the decay width corresponding to the di-photon decay of a general $\stateinequ{1}{S}{0}$ heavy quarkonium at order $v^2$, where $v$ denotes the relative velocity of the heavy quarks in the heavy quarkonium is given by~\cite{Guo:2011tz, Bodwin:2002cfe}:
	\begin{eqnarray} \label{final1}
	\Gamma(H(\stateinequ{1}{S}{0}) \rightarrow \gamma \gamma)&=&
	\frac{F_{\gamma\gamma}(\stateinequ{1}{S}{0})}{m_Q^2}\langle
	\mathcal{O}(\stateinequ{1}{S}{0}) \rangle_H
	+\frac{G_{\gamma\gamma}(\stateinequ{1}{S}{0})}{m_Q^4}\langle \mathcal{P}(\stateinequ{1}{S}{0}) \rangle_H \nn \\
      && +\frac{H^1_{\gamma\gamma}(\stateinequ{1}{S}{0})}{m_Q^6
      }\langle \mathcal{Q}^1_1(\stateinequ{1}{S}{0}) \rangle_H +\frac{H^2_{\gamma\gamma}(\stateinequ{1}{S}{0})}{m_Q^6
      }\langle \mathcal{Q}^2_1(\stateinequ{1}{S}{0}) \rangle_H,
	\end{eqnarray}
	where 
	\begin{align}
	\langle\mathcal{P}(\stateinequ{1}{S}{0})\rangle_{\textrm{LO}}=
	\mathbf{q}^2 \langle\mathcal{O}(\stateinequ{1}{S}{0})\rangle_{\textrm{LO}}.
	\end{align}
	The leading order long distance matrix element (LDME) is related to the wave function at the origin as
	\begin{eqnarray} \label{wfodefinition}
	\langle \mathcal{O}(\stateinequ{1}{S}{0})
	\rangle_H&=& 2 N_c |\psi(0)_{H}|^2.
	\end{eqnarray}
	Here $\mathbf{q}$ is half the relative three-momentum of the heavy quark and anti-quark that the heavy quarkonium consists of. $F_{\gamma\gamma}(\stateinequ{1}{S}{0})$, $G_{\gamma\gamma}(\stateinequ{1}{S}{0})$, $H^1_{\gamma\gamma}(\stateinequ{1}{S}{0})$, and $H^2_{\gamma\gamma}(\stateinequ{1}{S}{0})$ are the short distance coefficients obtained by equating the expression for the decay width of a heavy quarkonium decaying into light hadrons computed in perturbative QCD to that computed in perturbative NRQCD and are given by
	\begin{align}
	F_{\gamma \gamma}(\stateinequ{1}{S}{0})=& 2\pi \alpha^2 e_Q^4 \bigl(1+\frac{\alpha_s}{\pi} \frac{\pi^2-20}{3}\bigr)
	, \\
	G_{\gamma \gamma}(\stateinequ{1}{S}{0})=& 2\pi \alpha^2 e_Q^4
	[-\frac{4}{3}+
	\frac{\alpha_s}{\pi}\frac{1}{27}\bigl(48\ln\frac{\mu_{\Lambda}^2}{m_Q^2}-96\ln 2-15\pi^2+196\bigr)
	],\\
      H^1_{\gamma\gamma}+H^2_{\gamma\gamma}=&\frac{136 \pi}{45} e_Q^4 \alpha^2
	\end{align}
	
	\item \underline{$J/\psi\rightarrow e^+ e^- $}:\\
	The decay rate for $J/\psi$ decaying into an electron-positron pair is given to the first order in perturbative QCD corrections by~\cite{Bodwin:2002cfe, Lee:2018aoz}:
	\begin{eqnarray}
	\Gamma[ J/\psi \to e^+ e^-] &=&
	\frac{2 e_c^2 \pi \alpha^2}{ 3} \,
	\frac{\langle O_1 \rangle_{J/\psi} }{ m_c^2}\times \nn\\ 
 &&\left[1 - \frac{8}{3} \, \frac{\alpha_s}{ \pi}-\frac{1}{6} \langle v^2 \rangle_{J/\psi}+\frac{\alpha_s}{3 \pi}\bigg(\frac{8}{9}+\frac{8}{3} \ln\frac{\mu_{\Lambda}^2}{m_Q^2}\bigg)\langle v^2 \rangle_{J/\psi}+\frac{29}{18}\langle v^4 \rangle_{J/\psi}\right]^2. \ \ ~
	\label{gam-psi}
	\end{eqnarray}
\end{itemize}
The corresponding experimental limits for the BR's specified above are taken from PDG~\cite{Zyla:2020zbs} and are summarized in table~\ref{tab:radexptinptSwave}. The decay rates mentioned above are proportional to the square of the decay constants $f_{J/\psi}$ and $f_{\eta_c}$ which we have defined in eqs.~\ref{eq:fpsi} and \ref{eq:fetac}, respectively.

\begin{table*}[t]
	\begin{center}
 \resizebox{0.45\textwidth}{!}{
		\begin{tabular}{|c|c|c|c|c|}
			\hline
			\textbf{Mode}l& 
			\textbf{Branching Ratio (BR)~\cite{Zyla:2020zbs}}
			\\
			\hline
                $\eta_{c}\rightarrow\gamma\gamma$
			& $1.61(12)\times{10^{-4}}$ \\
			$J/\psi\rightarrow e^+e^-$
			& $5.971(32)\times{10^{-2}}$ \\
			\hline
		\end{tabular}
  }
		\caption{Experimental results for the modes used in our analysis. }
		\label{tab:radexptinptSwave}
	\end{center}
\end{table*}
A combined study of the charmonium rates mentioned above along with the lattice inputs on the decay constants and the form factors (at $q^2=0$) will be helpful to constrain $\psi^R_{B_c}(0)$, $\psi^R_{J/\psi}(0)$ and $\psi^R_{\eta_c}(0)$ simultaneously, which we will discuss in the following subsection.


\subsection{Analysis and results}

In this subsection, we discuss the numerical method to extract the radial wave functions $\psi^{R}_{B_c}(0)$, $\psi^{R}_{J/\psi}(0)$ and $\psi^{R}_{\eta_c}(0)$ using inputs on the form factors, the decay constants $f_{J/\psi}$, $f_{\eta_c}$ and $f_{B_c}$ as well as experimental data on radiative and rare decays of $J/\psi$, $\eta_c$ mentioned in subsection \ref{subsec:inputs}. The $B_c \to J/\psi $ and $B_c \to \eta_c$ form factors are sensitive to the products $\psi^R_{B_c}(0) \psi^R_{J/\psi}(0)$ and $\psi^R_{B_c}(0) \psi^R_{\eta_c}(0)$, respectively. The lattice inputs on $B_c\to J/\psi$ form factors and the input we have obtained on $B_c\to \eta_c$ form factors (eq.~\ref{eq:fbcetacq20}) at $q^2=0$ will hence be helpful in extracting these products of wave functions. On the other hand, the charmonium decay rate $\Gamma(J/\psi\to e^+e^-)$ and the decay constant $f_{J/\psi}$ are sensitive to $\psi^R_{J/\psi}(0)$. Similarly, $\Gamma(\eta_c \to \gamma\gamma)$ and $f_{\eta_c}$ are sensitive to $\psi^R_{\eta_c}(0)$. In principle, it is possible to constrain $\psi^R_{J/\psi}(0)$ and $\psi^R_{\eta_c}(0)$ directly from the experimental data given in Table~\ref{tab:radexptinptSwave} and from the lattice inputs on $f_{J/\psi}$ and $f_{\eta_c}$, respectively.

For the analysis, we follow a $\chi^2$-optimization technique. We construct a likelihood function comprising of the radial wave functions as parameters. These parameters enter into the likelihood function via the expressions for the observables BR($J/\psi\to e^+ e^-$), BR($\eta_{c}\to \gamma\gamma$), $f_{J/\psi}$, $f_{\eta_c}$ and the expressions for the $B_c\to J/\psi$ form factors at $q^2$=0 in NRQCD. The $\chi^2$ function corresponding to this likelihood is defined as
\begin{equation}\label{eq:chi22nd}
\chi^2= \sum_{i}\frac{(O_i^{NRQCD}-O_i^{expt})^2}{\sigma_i^2} + \sum_{i,j}(O_i^{NRQCD}-O_i^{Lattice})^T V^{-1}_{ij} (O_j^{NRQCD}-O_j^{Lattice}).
\end{equation}
Here, $O_i^{NRQCD}$'s represent the theory expressions for the form factors, decay constants and the decay rates calculated in the NRQCD effective theory. $O_i^{expt}$ and $O_i^{Lattice}$ represent the relevant experimental data and the lattice inputs, respectively. The $\sigma_i$ in the denominator of the first term represents the error in the experimental measurements at 1$\sigma$. Note that here, we use experimental data which are uncorrelated. Also, the lattice inputs on the decay constants are uncorrelated.

\begin{table}[t]
	\renewcommand{\arraystretch}{1}
	\centering
	\setlength\tabcolsep{5pt}
	\begin{tabular}{ccccccc}
		\hline
		\hline & $\alpha_{s}({\mu=m_c})$ & $\alpha_{s}({\mu=2 m_c})$   &$\alpha_{s}({\mu=m_b})$  &$m_b$ &$m_c$&Reference  \\
		&&&(GeV)&(GeV)&\\
		\hline Pole mass Scheme& $0.305(7)$ & 0.225(2)   & $0.2172(9)$&4.78&1.67& \cite{PhysRevLett.114.061802}\\
		$\overline{MS}$ Scheme&$0.359(4)$ & 0.250(2)  & $0.2271(6)$& 4.18&1.27&\cite{ParticleDataGroup:2020ssz}\\
		Kinetic Scheme&$0.401(6)$& 0.268(2)  & $0.2206(3)$&4.56&1.09&\cite{PhysRevLett.114.061802}\\
		\hline
		\hline
	\end{tabular}
	\caption{The values of the strong coupling constant $\alpha_{s}(\mu)$ (corrected at the 5-loop level~\cite{ParticleDataGroup:2020ssz}).}
	\label{tab:massscheme}
\end{table}

It is essential to realise that the form factors defined in NRQCD are susceptible to the masses of the $b$ and $c$ quarks. Therefore, it is expected that the extracted values of the radial wave functions will be highly scheme-dependent in general. For phenomenological analyses however, a single estimate (including errors) for each of these radial wave functions will be useful. Therefore, we will present the results in a scheme-independent way and estimate the error on the radial wave functions due to errors on the quark masses and the strong couplings that account for the scheme dependencies. In the fit, we have incorporated the values of $m_b$, $m_c$, and $\alpha_s$, which we obtain from the average of the values in the three schemes given in Table \ref{tab:massscheme} and added errors to take into account the ranges of values across the three schemes. Following this method, we have obtained
\begin{align}\label{eq:massavgscheme}
m_b &= 4.51 \pm 0.36, \ \ \ \ \ \ \ \ \ \ \ \ \ \ \ \ m_c = 1.34 \pm 0.27, \nonumber \\
\alpha_s(m_b) & =0.221 \pm 0.007, \ \ \ \ \ \  \alpha_s(2 m_c) =0.245 \pm 0.025. 
\end{align}
Note that the quark masses $m_b$ and $m_c$ have large errors due to the scheme dependence. We use these inputs as nuisance parameters in our fit.

Using the inputs for $m_b$, $m_c$ and $\alpha_s$ given in eq.~\ref{eq:massavgscheme}, we obtain the relative sizes of the NLO QCD ($\alpha_s$) and the relativistic corrections ($v^2$) with respect to the LO result in $f_{B_c}$ and in the form factors, which we have shown in eqs. \ref{eq:nloRCdecay} and \ref{eq:nloRCavgscheme}, respectively, in the appendix. The quoted 1$\sigma$ errors are due to the scheme dependence of $m_b$, $m_c$ and $\alpha_s$. Given the 1$\sigma$ error bars in the form factors, the NLO corrections could be as large as 35\%-40\% while the relativistic corrections could be as large as 50\%-60\%. Also, both corrections contribute with the same sign (positive) and have constructive interference with the LO results. We require large negative contributions from the uncomputed higher order corrections, like at order $\alpha_s^2$, $\alpha_s v^2$, $v^4$ and higher for the convergence of the perturbative series. In $f_{B_c}$, both the NLO and relativistic corrections have relative sign differences with respect to the LO contribution; the NLO correction could be as large as 10\% while the relative contribution from the LO relativistic correction is $< 5$\%, which is a small effect compared to those in the form factors.  
 
As we have mentioned earlier, the estimates of the higher-order corrections of the form factors are only partially known. The higher-order relativistic corrections of order $\mathcal{O}(v^4)$ or higher are not known. Furthermore, the perturbatively calculable QCD corrections beyond order $\alpha_s$ are unavailable. Similarly, for $f_{B_c}$ only the perturbative corrections at order $\alpha_s$ and the relativistic corrections at order $v^2$ are known. The uncomputed higher-order pieces in these form factors and $f_{B_c}$ could impact the extracted value of $\psi^R_{B_c}$(0). On the other hand, the corrections at order $\alpha_s$, $v^2$, $v^4$ and $\alpha_s v^2$ are known in the NRQCD estimates of $f_{J/\psi}$, $f_{\eta_c}$ which are relevant for the decay rates $\Gamma(J/\psi \to e^+e^-)$ and $\Gamma(\eta_c\to \gamma\gamma)$. Still, minor higher-order corrections that have been missing so far could influence the extracted values of the respective radial wave functions. Therefore, to take into account these effects numerically, we parametrise the missing corrections in the NRQCD expressions of the decay constants and form factors by unknown parameters $\delta_i$'s where `$i$' s indicate the respective form factor or the decay constant. We have extracted the $\delta_i$ simultaneously with the radial wave functions from a fit to all the given lattice inputs on the form factors, lattice estimates for the decay constants $f_{B_c}$, $f_{J/\psi}$, $f_{\eta_c}$ and the data on the measured branching ratios BR($J/\psi\to e^+ e^-$), BR($\eta_{c}\to \gamma\gamma$). In the fit, we have treated $\delta_i$s as nuisance parameters abiding by the Gaussian distribution function with mean value $0$ and 1$\sigma$ uncertainties of $0.4$ and $0.2$ in the form factors and $f_{B_c}$, respectively. At the same time, for the $\delta_i$s in $f_{J/\psi}$ and $f_{\eta_c}$ we consider the 1$\sigma$ error as $0.1$.

\begin{table}[t]
\centering
\resizebox{0.65\textwidth}{!}{
\begin{tabular}{|c|c||c|c|}
\hline
\multicolumn{2}{|c||}{\bf Fit results }& \multicolumn{2}{c|}{\bf Predictions using the fit results}\\
\hline
\textbf{Parameters} & \textbf{Fit results}& \textbf{Form Factor} & \textbf{Prediction}   \\
&& \& &\\
&& \textbf{decay constant} &\\
\hline
$\psi^R_{B_c}(0)$ & 0.916(124)&\text{$V^{J/\psi}(0)$}& \text{0.734(65)} \\
$\psi^R_{J/\psi}(0)$ & 0.835(40) &\text{$A_0^{J/\psi}(0)$}& \text{0.487(31)}  \\
$\psi^R_{\eta_c}(0)$ & 1.028(26)&\text{$A_1^{J/\psi}(0)$}& \text{0.462(28)} \\
$\delta_{V}$ & \text{-0.22(24)}&\text{$A_2^{J/\psi}(0)$}& \text{0.414(54)} \\
$\delta_{A_0}$&\text{-0.19(21)}&\textbf{$f^{\eta_c}_{+}(0)=f^{\eta_c}_{0}(0)$}& \text{0.64(15)} \\
$\delta_{A_1}$ & \text{-0.40(20)}  &\textbf{$f_{B_c}$}& \text{429(15)} \\
$\delta_{f_{+}}$& $-0.40(30)$&\text{$f_{J/\psi}$}& \text{404(6)} \\
$\delta_{f_{B_c}}$& $0.33(16)$ &\text{$f_{\eta_c}$}& \text{395(2)} \\
$\delta_{J/\psi ee}$ & $0.001(59)$ & & \\
$\delta_{\eta_c \gamma \gamma}$ & $-0.101(26)$ & & \\
\hline
\text{dof} & 8&&\\
\text{p-Value} & 0.08&&\\
\hline
\end{tabular}
}
\caption{The first two columns represent the numerical estimates for the radial wave functions extracted from a fit to the lattice inputs on $B_c \to J/\psi$ form factors at $q^2=0$, $B_c \to \eta_c$ form factors at $q^2=0$,  $BR(J/\psi \to e^+e^-)$, $BR(\eta_c\to \gamma\gamma)$ and the decay constants $f_{J/\psi}$, $f_{\eta_c}$, $f_{B_c}$, respectively. We have also considered additional errors ($\delta_i$) due to missing higher-order corrections as nuisance parameters. We present the predicted values of the form factors and decay constants using the fit results in the third and fourth columns.}
\label{tab:fitresultall}
\end{table}

We obtain the fit results after minimizing the $\chi^2$ function defined in eq.~\ref{eq:chi22nd}. The best-fit values and the respective 1$\sigma$ error for the radial wave functions and the $\delta_i$s which we have used as nuisance parameters in this fit are shown in table \ref{tab:fitresultall}\footnote{As we have mentioned, the available perturbative and relativistic corrections in the NRQCD form factors are large. To check their impact on the extracted values of $\psi^R_{B_c}(0)$, we have done separate fit incorporating, respectively, the NRQCD form factors at LO, LO+NLO, and LO+NLO+RC. We present the corresponding result in table~\ref{tab:fitresultsNRQCDWF} in the appendix. Please see the follow up discussion below table \ref{tab:fitresultsNRQCDWF} in the appendix.}. The radial wave functions $\psi^R_{J/\psi}(0)$ and $\psi^R_{\eta_c}(0)$ are bounded directly by the charmonium decay rates and the decay constants. Hence, the inputs on the form factors and $f_{B_c}$ are playing the main role in constraining $\psi^R_{B_c}(0)$. The estimated errors in $\psi^R_{B_c}(0)$ is about 15\% which are much larger then the respective errors in $\psi^R_{J/\psi}(0)$ and $\psi^R_{\eta_c}(0)$ which are $< 5$\%. This is not surprising since the major source of error in $\psi^R_{B_c}(0)$ is from scheme dependencies, which is about 80\% of the total error. We have mentioned earlier that the scheme dependencies are much larger in the NRQCD estimates of the form factors as compared to those in the decay constants $f_{J/\psi}$ and $f_{\eta_c}$, respectively. The dependence of the decay constants on $m_c$ is relatively weak. Therefore, even though $m_c$ has a significant error due to the scheme dependence, the extracted values $\psi^R_{J/\psi}(0)$ and $\psi^R_{\eta_c}(0)$ have a relatively small error due to this dependence.

The estimated $\delta_i$s associated with the form factors and $f_{B_c}$ have large errors, 80\% of which can be attributed to the scheme dependencies of the form factors and $f_{B_c}$. To simultaneously explain the given inputs on the form factors and $f_{B_c}$, the additional uncomputed errors in the form factors could be large and destructively interfere with the respective leading order known contributions. As we have mentioned earlier, this is also a requirement for the convergence of the perturbative series. In addition, the estimated size of these uncomputed corrections is smaller than the respective leading order contributions. Similarly, the uncomputed errors in $f_{B_c}$ could be as large as 30\% and interfere constructively with the known leading order contribution. However, this additional contribution has the opposite sign compared to the known corrections at order NLO-QCD and $v^2$. The overall impact of the uncomputed errors in the decay constants $f_{J/\psi}$ and $f_{\eta_c}$ are relatively small $< 10\%$. Using these fit results, we have predicted the $B_c \to J/\psi$ and $B_c \to \eta_c$ form factors and the three decay constants which we have shown in the third and fourth columns of the same table. Note that all the form factors and decay constants are consistent with the respective lattice inputs at the 1$\sigma$ CI in this fit. We will use the fit results of table \ref{tab:fitresultall} to predict the observables discussed in the following sections.

The extracted radial wave-function estimates are used to predict and update the production and decay channels involving these charmonia under the NRQCD framework which we will discuss in the upcoming sections.

\section{Non-leptonic decays of $B_c$ mesons into charmonium states}\label{sec:non-lep}
Non-leptonic decays of heavy mesons offer an opportunity to understand the nature of Quantum Chromodynamics(QCD). The $B_c$ meson can decay non-leptonically via the decays of the $\bar{b}$ quark with the c quark as the spectator or vice versa. There could also be annihilation dominated decays of $B_c$ mesons. We study the exclusive non-leptonic two-body $B_c$ decays within the factorization approximation with the leading order non-factorizable corrections calculated in NRQCD \cite{Qiao:2012hp,Zhu:2017lqu,Zhu:2017lwi}. As a leading order approximation, naive factorization is widely used to study these decays in which the non-leptonic decay amplitudes reduce to the product of the form factor and a decay constant. For the decays we have considered in this analysis, the amplitudes are expressed as the product of the decay constant and the transition form factors for $B_c \to J/\psi (\eta_c)$ decays. For the decay constant, we have used the corresponding lattice estimates (tab.~\ref{tab:Gigenmoments}) and the lattice estimates for the relevant form factors in $B_c\to J/\psi$ decays (tab.~\ref{tab:inputlattQCDSR}). At the same time, for $f_{+,0}^{B_c\to \eta_c}$ we use our estimates given in the section \ref{sec:Bctoetac}. In evaluating the non-factorizable corrections, we have used our estimates for radial wave functions given in table \ref{tab:fitresultall}. The branching fractions of a few such non-leptonic decays have been measured.

\begin{figure}[h!]
	\centering
	\begin{tikzpicture}[baseline={(current bounding box.center)}]
	\begin{feynman}
	\vertex(a);
	\vertex[right=2cm of a](a1);
	\node[right=2cm of a,  dot, style=black];
	\node[above right=1cm and 1.55cm of a1, dot, style=black];
	\vertex[right= 2cm of a1] (a2);
	\vertex[above right=1cm and 1.55cm of a1] (a4);
	\vertex[above right=1cm and 2.6cm of a1] (a5);           
	\vertex[above right=2cm and 2cm of a1] (a6);           
	\vertex[right=0.75cm of a] (a3);
	\vertex[below=2cm of a] (b);
	\vertex[right=4cm of b] (c1);
	\vertex[right=0.75cm of b] (c2);
	\diagram* {
		(a) -- [anti fermion ,arrow size=1pt,edge label={\(\overline{b}\)}] (a1) -- [anti fermion ,arrow size=1pt,edge label={\(\overline{c}\)}] (a2) ,(b) --[fermion ,arrow size=1pt,edge label'={\(c\)}] (c1),(a)--[fill=gray!15,plain,bend left](b),(a)--[fill=gray!15,plain,bend right,edge label'={\(B_c^+\)}](b),(a2)--[fill=gray!15,plain,bend left,edge label={\((c\overline{c})\)}](c1),(a2)--[fill=gray!15,plain,bend right](c1), (a4)[dot] --[boson ,arrow size=1pt, edge label'={\(\rm W^+\)}] (a1),(a4) [dot] --[anti fermion ,arrow size=1pt, edge label'={\(\rm \overline{q}\)}] (a5),(a6) --[anti fermion ,arrow size=1pt, edge label'={\(\rm u\)}] (a4),(a5)--[fill=gray!15,plain,bend left](a6),(a5)--[fill=gray!15,plain,bend right,edge label'={\(P,V\)}](a6),};
	\end{feynman}
	\end{tikzpicture} 
	\caption{Tree level Feynman diagrams for the non-leptonic $B_c \to (c\bar{c})(P, V)$ decays, where P and V stand for a light pseudo-scalar meson and a vector meson and (c$\bar{c}$) stands for S wave charmonium.}
	\label{fig:feynnonlep}
\end{figure} 

\subsection{Formalism}
The theoretical description of the non-leptonic decays involves the matrix elements of the local four-fermion operators. The effective and CKM favored Hamiltonian for the $b\to c \bar{u}d$ transition can be written as follows:
\begin{equation}\label{eq:heffnl}
\mathcal{H}_{eff}=\frac{G_F}{\sqrt{2}}V^*_{ud} V_{cb} \big(C_1 (\mu) Q_1(\mu)+C_2(\mu) Q_2({\mu})\big),
\end{equation}
Here, $V_{ud}$, $V_{cb}$ are the CKM matrix elements, and the $C_i({\mu})$s are the Wilson coefficients (WCs) which take into account the short-distance effects. The WCs $C_{1,2}(\mu)$ are evaluated perturbatively at the W scale and are then evolved down to the renormalization scale $\mu \approx m_b$ by the renormalization group equations. The effects of soft gluons below the scale $\mu$ with the virtualities remain in the hadronic matrix elements of the local four-fermion operators $Q_i$. The four-fermion effective operators $Q_{1,2}(\mu)$ are defined as
\begin{align}
Q_1 &= \bar{d}_{\alpha}\gamma^{\mu} (1-\gamma_{5})u_{\alpha} \bar{c}_{\beta}\gamma_\mu(1-\gamma_{5})b_{\beta}, \nonumber \\
Q_2 &= \bar{d}_{\alpha}\gamma^{\mu} (1-\gamma_{5})u_{\beta} \bar{c}_{\beta}\gamma_\mu(1-\gamma_{5})b_{\alpha}, 
\end{align}
where $\alpha$ and $\beta$ are color indices, and the summation conventions over repeated indices are understood. The Fierz rearrangement
\begin{eqnarray}
T^{A}_{\alpha \beta} T^{A}_{\rho \lambda}=-\frac{1}{6} \delta_{\alpha \beta} \delta_{\rho \lambda}+\frac{1}{2} \delta_{\alpha \lambda} \delta_{\rho \beta},
\end{eqnarray} 
can be used to change the above basis to 
\begin{eqnarray}
Q_0=Q_1, ~~~~Q_8=-\frac{1}{6}Q_1+\frac{1}{2}Q_2,
\end{eqnarray}
with the following Wilson coefficients (WCs)  \cite{Buchalla:1995vs}
\begin{eqnarray}
C_{0}=C_1+C_2/3= \frac{2}{3}C_{+}+\frac{1}{3}C_{-}\,,~~C_{8}=2 C_2 = C_{+}-C_{-}\,,
\end{eqnarray}
where
\begin{eqnarray}
C_{\pm}=\left[\frac{\alpha_s(M_W)}{\alpha_s(\mu)}\right]^{
	\frac{\gamma_\pm} {2\beta_0}}\,,~~\gamma_\pm=\pm 6\frac{N_c\mp
	1}{N_c}\,,~~\beta_0=\frac{11N_c-2n_f}{3}\,.
\end{eqnarray}
In this paper, we have only focused on the CKM favored processes. Here, we do not consider non-leptonic $B_c$ decays into $D^{(*)},$ $D_s^{(*)}$ mesons as these decays are strongly CKM suppressed. We limit our analysis of the $B_c$ non-leptonic decays to the case when the final meson $M_{c\bar{c}}$ is charmonium, and the light $M$ meson is $\pi$, $K^{(*)}$, $\rho$. The corresponding Feynman diagram is shown in fig~\ref{fig:feynnonlep}.

\begin{table}[t]
\renewcommand{\arraystretch}{1}
\centering
\setlength\tabcolsep{6pt}
\label{}
\begin{tabular}{ccc|c|c|}
\hline
\multicolumn{3}{c|}{Gegenbauer Coefficients} & Decay constant & Form factors \\
\cline{1-3}
Mesons& Coefficient & Values &  (MeV) \cite{ParticleDataGroup:2020ssz}  &    \\
\hline
$\pi$~\cite{RQCD:2019osh}&$a_1$&$0$ & $f_{\pi} = 130.5$ &  $f_0(m_{\pi}^2) = 0.45(20)$   \\
&$a_2$&$0.116^{+19}_{-20}$ &  & \\
\hline
$K$~\cite{RQCD:2019osh}& $a_1$&$0.0525^{+31}_{-33}$ & $f_{K} = 155.72$ & $f_0(m_K^2)= 0.45(20)$  \\
& $a_2$&$0.106^{+15}_{-16}$ & &  \\
\hline
&$a_1^{||,\perp}$&0 &  &    \\
$\rho$~\cite{Braun:2016wnx}&$a_2^{||}$&$0.132 \pm 0.027$ & $f_{\rho} = 221$ & $f_+(m_{\rho}^2) = 0.47(20)$   \\
&$a_2^{\perp}$&$0.101\pm 0.022$ &  &  \\
\hline
&$a_1^{||}$&$0.03\pm 0.02$ &  & \\
$K^*$~\cite{Ball:2007rt}&$a_1^{\perp}$ & $0.04\pm0.03$ & $f_{K^{*}} = 220$ & $f_+(m_{K^*}^2)  = 0.48(20) $ \\
&$a_2^{||}$&$0.11 \pm 0.09$ & &  \\
&$a_2^{\perp}$&$0.10\pm 0.08$ & & \\
\hline
\hline
\end{tabular}
\caption{Gegenbauer moments for the twist-2 distribution amplitudes of light mesons and the inputs for the decay constants used in the estimate of the branching fractions of the non-leptonic decays of $B_c$ meson.}
\label{tab:Gigenmoments}
\end{table}

Following the naive-factorization approximation the amplitude can be expressed as
\begin{equation}
\langle M_{c\bar{c}} M| \mathcal{H}_{eff}|B_c\rangle \propto C_i \langle M_{c\bar{c}}| (\bar{b}c)_{V-A}|B_c\rangle \times \langle M| (\bar{q_1} q_2)_{V-A}|0\rangle \approx C_i f_M F^{B_c \to M_{c\bar{c}}},
\end{equation}
where $f_M$ and $F^{B_c \to M_{c\bar{c}}}$ are the decay constant and the form factor, respectively. Note that in the above equation, the matrix elements of the weak current between the vacuum and a pseudoscalar (P) or a vector (V)  meson have been parametrized by the decay constant $(f_P,f_V)$ and defined as:
\begin{equation*}
\langle P(p_{\mu})|{(\bar{q_1} q_2)}_{V-A}|0\rangle =- i f_P p_{\mu},~~~\\\nonumber
\langle V|{(\bar{q_1} q_2)}_{V-A}|0\rangle =- i f_V m_V \epsilon^*_{\mu},
\end{equation*}
where  $m_V$ and $\epsilon_{\mu}$ are the mass and polarization vectors of the vector meson. The matrix elements $\langle M_{c\bar{c}}| (\bar{b}c)_{V-A}|B_c\rangle$ are already defined in eq.~\ref{eq:qcdffdef}. 
Note that, in general the matrix element between the vacuum and the mesonic state can be parametrized by the mesonic wave function and its decay constant as follows:
\begin{align}\label{eq:decaycons}
\langle 0|q_1(0) \gamma^{\mu} \gamma_5 q_2(0)|P(P)\rangle &=if_P P^{\mu}\int_{0}^{1} dx \phi_P(x),\\
\langle 0|q_1(0) \gamma^{\mu}  q_2(0)|V(P),\epsilon_{\lambda=0}\rangle &=if_V M_V\epsilon^{\mu} \int_{0}^{1} dx \phi_{V_{\parallel}}(x),\\
\langle 0|q_1(0) \sigma^{\mu \nu}  q_2(0)|V(P),\epsilon_{\lambda=\pm 1}\rangle &=if_V^{\perp} \int_{0}^{1} dx (\epsilon^{\mu} P^{\nu}-\epsilon^{\nu} P^{\mu})~ \phi_{V_{\perp}}(x).
\end{align}	
Hence, following the QCD factorization approach, the most general expression for the decay amplitude for $B_c \to M_1(c\bar{c}) M_2$ decay is given by 
\begin{equation}\label{eq:QCDF}
{\cal A}(B_c \to M_1(c\bar{c}) M_2 ) \sim F^{B \to M_1}(q^2=M_2^2) f_{M_2} \int_{0}^{1}{dx\ T_H(x)\ \phi_{M_2}}(x) + \mathcal{O}(\frac{1}{m_b}) + \mathcal{O}(v^2),
\end{equation} 
where $F^{B \to M_1}$ is the form factor for $B \to M_1$ and $f_{M_2}$ is the decay constant corresponding to $M_2$ meson. $M_1$ represents the meson that takes away the spectator quark. $\phi_{M_2}$ is the light-cone-distribution amplitude (LCDA) of the meson $M_2$, and the $T_{H}$ are the perturbatively calculable hard-scattering kernels for the various operators in the effective weak Hamiltonian (eq.~\ref{eq:heffnl} ). 
In the naive factorization approximation, $T_H$ is independent of $x$, and $\int_{0}^{1}{dx\ \phi_{M_2}}(x) = 1$ is obtained from the normalization. Hence, under this approximation, the amplitude is proportional to the product of the decay constant and the form factor. 
Once we add the non-factorizable contributions, i.e. contributions from the diagrams containing gluon exchanges that do not belong to the form factor for the $B_c\to M_1$ transition, the hard scattering kernels given in eq.~\ref{eq:QCDF} can be generalized as \cite{Qiao:2012hp},

\begin{equation}
{\cal A}(B_{c}^{+}\to J/\psi(\eta_{c}) P)=
\frac{G_{F}}{\sqrt{2}}V_{ud}^{*}V_{cb} \int_{0}^{1}{dx\ \phi_P(x)\Big[C_0(\mu)   
	\big(T_{f,0}(\mu) + T_{nf,0}(x,\mu)\big) \langle Q_{0} \rangle
	+ C_{8}(\mu)   T_{nf,8}(x,\mu) \langle Q_{8}\rangle  \Big]}\label{eq:NLamplitude},
\end{equation}

\begin{equation}
T_{f,i}(\mu)=\sum_{k=0}^\infty (\frac{\alpha_s}{4\pi})^k
T^{(k)}_{f,i}(\mu)\,,~~~T_{nf,i}(\mu)=\sum_{k=0}^\infty (\frac{\alpha_s}{4\pi})^k
T^{(k)}_{nf,i}(\mu)\,,
\end{equation}
where $P$ is a pseudoscalar meson. In these equations, the hard kernels $T_{f}$ and $T_{nf}$ represent the factorizable and non-factorizable contributions in the decay amplitudes, respectively, which are perturbatively calculable. These can further be classified as the contributions to $\langle Q_0 \rangle$ and $\langle Q_8 \rangle$, respectively. In NRQCD, the matrix elements $\langle Q_{0} \rangle$ and $\langle Q_{8} \rangle$ for $B_c \to \eta_c$ decays can be written as \cite{Qiao:2012hp}:

\begin{align}\label{eq:matrixBctoetac}
\langle Q_0 \rangle^{B_c\to \eta_c P}=& (M_{B_c}^2 - M_{\eta_c}^2) f_P f^{B_c\to \eta_c}_0(q^2 = M_P^2)           \\  \nonumber 
\langle Q_8 \rangle^{B_c\to \eta_c P}=&\frac{2\sqrt{2} \pi f_{P} \psi_{\eta_c}(0)\psi_{B_c}(0)  C_A C_F \alpha_s \sqrt{m_b+ m_c}(m_b+3m_c)^2(x m_c-(x-1)m_b)}{m_c^{3/2} N_c^2 (m_c-m_b)\big((x-1) m_b+(3x-2) m_c\big)\big(x m_b+(3x-1)m_c\big)}.
\end{align}

The same matrix elements for $B_c\to J/\psi$ decays are obtained as \cite{Qiao:2012hp}
\begin{align}\label{eq:matrixBctopsi}
\langle Q_0 \rangle^{B_c \to J/\psi P}=& - 2 m_{J/\psi} f_P P_{B_c}\cdot\epsilon^*_{J/\psi} A_0^{B_c\to \eta_c}(q^2 = M_P^2)  \\\nonumber
\langle Q_8 \rangle^{B_c \to J/\psi P}=-&\frac{8\sqrt{2} \pi f_{P} \psi_{J/\psi}(0)\psi_{B_c}(0) P_{B_c}\cdot\epsilon^*_{J/\psi} C_A C_F \alpha_s (m_b+ m_c)^2}{m_c^{1/2} N_c^2 (m_b-m_c)^2\big((x-1) m_b+(3x-2) m_c\big)\big(x m_b+(3x-1)m_c\big)}\times\\\nonumber
&\bigg(3(2x-1)m_b m_c +(x-1) m_b^2+(9x-4) m_c^2\bigg).
\end{align}

Here, $\psi_{c\bar{c}}(0)$ and $\psi_{M_2}(0)$ are the Schr{\"o}dinger wave function defined earlier. In the above equations, we define $P_{B_c}\cdot\epsilon^*_{J/\psi}=\frac{m_{B_c}}{m_{J/\psi}}|p|$, $|p|=\frac{\sqrt{[m_{B_c}^2-(m_{J/\psi}+m_{p})^2][m_{B_c}^2-(m_{J/\psi}-m_{p})^2]}}{2 m_{B_c}}$, $z= \frac{m_c}{m_b}$, $m_{B_c}=(m_b+m_c)$, $m_{J/\psi}= m_{\eta_c}=2m_c$. In the above expressions, we have used the inputs on the form factors extracted from  lattice as discussed earlier. Note that at the tree level, the hard kernel $T_{f,0}^0=1$ and independent of $x$. At the one-loop level $T_{f,0}^1(\mu)$ does not depend on $x$ either.  For these types of contributions the convolution integral in eq.~\ref{eq:NLamplitude} could simply be written as $\int_{0}^{1}{dx\ \phi_P(x)} = 1$, the normalization condition. At the lowest order for both the decay modes $T_{nf,0}^0=0$ and $T_{nf,8}^0 =1$. The detailed mathematical expressions for the $T_{f,0}^1$, $T_{nf,0}^1$ and $T_{nf,8}^1$ can be obtained from ref.\cite{Qiao:2012hp}. However, we have not taken into account the non-factorizable contributions $T_{nf,0}^1$ and $T_{nf,8}^1$ in our analysis. The LCDA of light meson can be written as an expansion in Gegenbauer polynomials, defined as follows:
\begin{equation}\label{eq:LCDApseudo}
\phi_{M} (x,\mu^2)=6x(1-x) \big[ 1+ \sum_{n=1}^{\infty} a_n^M (\mu^2) C_n^{3/2} (2x-1)\big],
\end{equation}
where the Gegenbauer polynomial is given by the coefficient $C_n^{3/2}(2x -1)$ and $a_n^{M}$s are the Gegenbauer moments. 
In the asymptotic limit, the above equation can be written as $\phi_M(x) \approx 6 x (1-x)$. 
Following the definitions given in eq.~\ref{eq:decaycons} we obtain that only the parallel component of the wave functions contributes to the decay amplitude in $B_c \to \eta_{c} V(\rho, K^*)$ decays. The transition amplitude for $B_c \to \eta_c V$ will have only the longitudinal components $\mathcal{A}_{||}$ which can be expressed as  
\begin{align*}\label{eq:NLamplitude}
\mathcal{A}_{||}(B_c \to \eta_c~ V)&=
\frac{G_{F}}{\sqrt{2}}V_{ud}^{*}V_{cb} \int_{0}^{1}dx\ \phi_{V,||}(x)\times\nn \\ &\biggl(C_0(\mu)   
	\big(T_{f,0}(\mu) + T_{nf,0}(x,\mu)\big) \langle Q_{0} \rangle^{B_c \to \eta_c V}
	 + C_{8}(\mu)   T_{nf,8}(x,\mu) \langle Q_{8}\rangle^{B_c \to \eta_c V}  \biggr),
\end{align*}
with the twist-2 LCDA given by 
\begin{equation}\label{eq:LCDAvector}
\phi_{V,||} (x,\mu)=6x(1-x) \big[ 1+ \sum_{n=1}^{\infty} a_{n,||}(\mu) C_n^{3/2} (2x-1)\big],
\end{equation}
The matrix elements $\langle Q_{0} \rangle^{B_c \to \eta_c V}$ and $\langle Q_{8}\rangle^{B_c \to \eta_c V}$ can be obtained from eq.~\ref{eq:matrixBctoetac} with the replacement $f_P \to f_V$ and $f_0(q^2) \to f_+(q^2)$. In this analysis, we will predict the branching fractions of the $B_c \to J/\psi (\pi,K)$ and $B_c \to \eta_c (\pi,K,K^*,\rho)$ decays. The available inputs on the Gegenbauer moments are given in table \ref{tab:Gigenmoments} which we have used in our analysis. 

\subsection{Results}

\begin{table}[t!]
\begin{center}
\resizebox{0.75\textwidth}{!}{
\begin{tabular}{|*{6}{c|}}
\hline
$\text{\bf Decay modes}$ &\multicolumn{2}{c|}{\bf Our predictions} &\multicolumn{3}{c|}{\text{\bf Other estimates}}  \\
\cline{2-6}
$(\times 10^{3})$ &  $\text{\bf Naive Fac.}$  & {\bf with  corrections} & \bf{PQCD} & \bf{RQM} & \bf{QCDSR}\\
&  &  {\bf (fac. + Non-fact. (LO))} &\cite{PhysRevD.90.114030} &\cite{PhysRevD.68.094020} & \cite{Kiselev:2002vz} \\
\hline
$\mathcal{B}(B_c\to \text{J/$\psi \pi $})$  &  $\text{0.702(92)}$    &  $\text{0.693(92)}$ &$2.33(63)$&$0.61$ &$1.3$  \\
$ \mathcal{B}(B_c\to\text{J/$\psi $K}) $  &  $\text{0.053(7)}$    &$\text{0.052(7)}$&$0.19(4)$& $0.05$&$0.11$ \\
$ \mathcal{B}(B_c\to \eta _c\pi)$  &  $\text{0.66(41)}$  &$\text{0.63(40)}$ &$2.98(84)$ & $0.85$ &$2.0$  \\
$ \mathcal{B}( B_c\to\eta _c K)$  &  $\text{0.051(31)}$    &  $\text{0.048(31)}$&$0.24(4)$&$0.07$&$0.13$  \\
$\mathcal{B}( B_c\to \eta_c\rho)$  &  $\text{1.9(11)}$    &$\text{1.8(11)}$ &$9.83(138)$ & $2.1$& $4.2$ \\
$ \mathcal{B}(B_c\to \eta_c K^*)$  &  $\text{0.100(59)}$   & $\text{0.095(58)}$ &$0.57(10)$ & $0.11$ & $0.20$   \\
\hline
\end{tabular}
}
\caption{The predictions for the branching fractions for a few $B_c \to \eta_{c}(P,V)$, and $B_c \to J/\psi P$ decays. For the predictions with non-factorizable corrections the radial wave functions are taken from tab.~\ref{tab:fitresultall}. Our predictions have been compared with those obtained in other methods/models.}
\label{tab:brnonleppredkin}
\end{center}
\end{table}

Following the formalism and the relevant inputs discussed above, we have predicted the respective branching fractions. We have presented the results in the naive factorization approximation and including non-factorizable corrections at LO (discussed above), respectively. As discussed before, for the numerical estimates we need the inputs on the form factors $A_0^{B_c\to J/\psi}(M^2)$ and $f_{0,+}^{B_c\to \eta_c}(M^2)$ where $M$ is the mass of the light meson to which $B_c$ is decaying. Following are the lattice inputs for $A_0$ and $f_0$: 
\begin{equation}\label{eq:A0input}
A_0^{B_c\to J/\psi}(m_{\pi}^2)  = 0.478 \pm 0.031 ,\ \ \ \ \ \ \ f_0^{B_c\to \eta_c}(m_{K}^2)=0.45 \pm 0.20
\end{equation}
which we have extracted from the results of ref.~\cite{Harrison:2020gvo}. The inputs for the $f_0$'s are shown in table \ref{tab:Gigenmoments}, which we have obtained from our analysis in section \ref{sec:Bctoetac}. For the numerical estimates of the non-factorizable corrections calculated in NRQCD, we need inputs on the radial wave functions which we have taken from table~\ref{tab:fitresultall}. The inputs on Gegenbauer coefficients for the light meson wave functions have been taken from table \ref{tab:Gigenmoments}. 
\begin{table}[t]
    \centering
    \resizebox{0.65\textwidth}{!}{
    \begin{tabular}{|*{4}{c|}}
    \hline
  $\text{Ratio of}$  &  \multicolumn{2}{c|}{\bf Our predictions}& \text{Expt. Obs.}\\
  \cline{2-3}
   \text{BR. fractions}&$\text{Naive Fact.}$  &  $\text{With correction}$ & \cite{Zyla:2020zbs} \\
       &  &  $\text{fac.+Non-fact. (LO)}$ & \\
  \hline
  $\frac{\mathcal{B}(\text{J/$\psi $}+K)}{\mathcal{B}(\text{J/$\psi $}+\pi \
)}$  &  $\text{0.0755(3)}$  &  $\text{0.0756(3)}$ & $0.079(8)$\\
  $\frac{\mathcal{B}(\text{$\eta_c$}+\rho )}{\mathcal{B}(\text{$\eta_c$}+\
\pi )}$  &  $\text{2.835(24)}$  &  $\text{2.838(25)}$ &\\
  $\frac{\mathcal{B}\left(\text{$\eta_c$}+K^*\right)}{\mathcal{B}(\text{$\eta_c$}+\pi )}$  &  $\text{0.149(2)}$  &  $\text{0.149(2)}$  \ & \\
  $\frac{\mathcal{B}(\text{$\eta_c$}+K)}{\mathcal{B}(\text{$\eta_c$}+\pi \
)}$  &  $\text{0.0764(3)}$  &  $\text{0.0764(3)}$ & \\
\hline
 \end{tabular}
 }
    \caption{Comparisons of the results of the ratios of the Branching fractions with the existing experimental measurements.}
    \label{tab:ratiosNLBF}
\end{table}
Using all the relevant inputs discussed above, we predict the respective branching fractions shown in table \ref{tab:brnonleppredkin}. The estimated errors in the branching fractions of $B_c \to J/\psi \pi (K)$ decays are $\approx$ 15\%. Furthermore, the estimated errors are the same in both the naive factorization and with the corrections (both the factorizable and non-factorizable) discussed above. However, the estimated errors in the non-leptonic decays involving the $\eta_c$ meson in the final state have large errors. The significant sources of error in these estimates are the form factors. The factorizable contributions depend only on the form factors and the decay constants. The non-factorizable corrections depend on the radial wave functions, and these corrections are $\lsim 0.5$\% of the factorizable contributions and, therefore, highly suppressed. The estimated large errors on the radial wave functions do not impact the estimated error on the respective non-leptonic branching fractions after incorporating the non-factorizable corrections. We have compared our results with other estimates based on QCD models. In PQCD \cite{PhysRevD.90.114030}  and QCDSR \cite{PhysRevD.68.094020} approaches, the estimated values are relatively higher due to the different inputs for the form factors used in both analyses. However, our predictions include the estimates presented in the relativistic quark model \cite{Kiselev:2002vz}, though they have not provided any error in their estimates. We have also estimated a couple of ratios of the branching fractions, which we have presented in table \ref{tab:ratiosNLBF}, which could be checked in future measurements. Note that the estimated errors in these ratios have reduced considerably due to the cancellations of $B_c\to J/\psi$ or the $B_c\to \eta_c$ form factors.


\section{Predictions for other production and decay modes for charmonium}
 Using the results from table~\ref{tab:fitresultall}, one could update the predictions of various decay rates and production or annihilation cross-sections calculated in the NRQCD effective theory involving these charmoniums. This is not the goal of this paper. However, we update the predictions of a few interesting channels in this section. 
\subsection{Electron positron annihilation to charmonium}
We have studied $e^+ e^-$ decays into double charmonium in the scope of the NRQCD framework. The corresponding results are presented below in table~\ref{tab:epemdoublechar}. The explicit expressions for the cross-sections corresponding to these decay channels are presented below.
\begin{itemize}
	\item \underline{$e^{+}e^{-}\rightarrow J/\psi +\eta_{c}$}:\\
	The production rate up to $\mathcal O(\alpha_s v^2)$ can be written as~\cite{Dong:2012xx,Li:2013otv}.
	\begin{equation}
	\sigma[e^+e^-\rightarrow J/\psi + \eta_c]=\sigma_0+ \sigma_2+\mathcal{O}(\sigma_0 v^4)
	\end{equation}
	Where,
	\begin{align}
	\sigma_0= \frac{8 \pi \alpha^2 m_c^2(1-4r)^{3/2}}{3}\langle \mathcal{O}_1\rangle _{J/\psi} \langle\mathcal{O}_1\rangle_{\eta_{c}}|c_0|^2
	\end{align}
	\begin{align}
	\sigma_2=& \frac{4 \pi \alpha^2 m_c^2(1-4r)^{3/2}}{3}\langle \mathcal{O}_1\rangle _{J/\psi} \langle\mathcal{O}_1\rangle_{\eta_{c}}\biggl\{\bigg(\frac{1-10r}{1-4r}|c_0|^2+4 Re[c_0c^*_{2,1}]\bigg)\langle v^2 \rangle_{J/\psi}\\ & +(\frac{1-10r}{1-4r}|c_0|^2+4 Re[c_0c^*_{2,2}]\bigg)\langle v^2 \rangle_{\eta_{c}}\biggr\}
	\end{align}
	where $c_0, c_{2,1}, c_{2,2}$ are given in ref~\cite{Dong:2012xx,Li:2013otv}. The measured values of this cross section by the BaBar collaboration is given by \cite{BaBar:2005nic}
	\begin{equation}
\sigma[e^+e^-\rightarrow J/\psi + \eta_c] \mathcal{B}^{\eta_c}[\ge 2] = 17.6 \pm 2.8^{+1.5}_{-2.1} ~\text{fb},
	\end{equation}
while that from the Belle collaboration \cite{Belle:2002tfa, Belle:2004abn} is 
	\begin{equation}
	\sigma[e^+e^-\rightarrow J/\psi + \eta_c]\mathcal{B}^{\eta_c}[\ge 2] = 25.6 \pm 2.8 (\text{stat.}) \pm 3.4 (\text{syst.})~\text{fb}.
	\end{equation}
\end{itemize}
In the above equations, $\mathcal{B}^{\eta_c}[\ge 2]$ denotes the branching fraction of $\eta_c$ decaying into at least two charged tracks. Hence, the right-hand side of the above equations represents the lower limits of the scattering cross sections $\sigma[e^+e^-\rightarrow J/\psi + \eta_c]$ obtained from the respective experiments.
\begin{table}[ht!]
	\begin{center}
		\begin{tabular}{|*{3}{c|}}
			\hline
			\textbf{Decay Channel}  & $\mu$ & $\textbf{Our Predictions}$ ($\sigma$ in fb) \\
			\hline
			$ \sigma(e^+e^-\to J/\psi \eta _c )$ & $2 m_c$  & $23\pm 7$ \\
			$\sigma(e^+e^-\to J/\psi \eta _c) $ &$\frac{\sqrt{s}}{2}$ & $20\pm 8$  \\
			$\sigma(e^+e^-\to \eta _c \gamma )$ &$\frac{\sqrt{s}}{2}$ & $56 \pm 15$\\
			\hline
		\end{tabular}
		\caption{Cross-section estimates for $e^+ e^-$ decay into charmonium. The radial wave functions are taken from table \ref{tab:fitresultall}.}
		\label{tab:epemdoublechar}
	\end{center}
\end{table}
\begin{itemize}
	\item \underline{$e^{+}e^{-}\rightarrow \eta_{c}\gamma $}
\end{itemize}
An electromagnetic transition between quarkonium states offers the distinctive experimental signature of a monochromatic photon, a useful production mechanism for the discovery and study of the lower-lying state, and a unique window on the dynamics of such systems. In this regard, we predict the scattering cross section $\sigma(e^+e^- \to \eta_c\gamma)$ using our result. The corresponding expression for the cross-section including the available corrections at order $\alpha_s$, $v^2$ and $\alpha_s v^2$ is given by ~\cite{Xu:2014zra}
\begin{equation}
\sigma=\hat{\sigma}^{(0)}_{\eta_c}\big[1 + \alpha_{s} c^{10} + (c^{02} + \alpha_{s} c^{12})\langle v^2\rangle\big] \langle0|\mathcal{O}^{\eta_c}|0\rangle,
\end{equation}
where the LO short-distance cross section for $\eta_c$ is given by
\begin{eqnarray}
\hat{\sigma}^{(0)}_{\eta_c}=\frac{(4\pi\alpha)^3Q_c^4(1-r)}{6\pi m_c s^2}.
\end{eqnarray}

The analytical expressions for the asymptotic behaviour of $c^{02}$, $c^{10}$ and $c^{12}$ at high energies ($r\to 0$) are taken from ref.~\cite{Xu:2014zra}, which are as follows
\begin{eqnarray} 	
\lim\limits_{r\rightarrow0}c^{02}&=&-\frac{5}{6}\nonumber\\	
\lim\limits_{r\rightarrow0}c^{10}&=&-\frac{2}{9\pi}[3(3-2\ln2)\ln r + 9(\ln^2 2-3\ln2 + 3) +\pi^2] \nonumber\\ 
&\approx&-0.34\ln r-1.59	\\ 
\lim\limits_{r\rightarrow0}c^{12}&=&\frac{1}{27\pi}[3(21-10\ln2)\ln r+15(3\ln^2 2-7\ln2)+28+5\pi^2] \nonumber\\
&\approx&0.50\ln r+0.31\nonumber	
\label{eqn:asymp}
\end{eqnarray}
The LO long-distance matrix elements are obtained from the radial wave functions at the origin:
\begin{eqnarray}
\langle0|\mathcal{O}^{\eta_c}|0\rangle&=&\frac{2N_c|\psi_{\eta_c}^R(0)|^2}{4\pi}
\end{eqnarray}

Note that the cross sections defined above are dependent on $\alpha_s$, $m_c$ and on the radial wave functions $\psi^R_{J/\psi}(0)$, $\psi^R_{\eta_c}(0)$, respectively. In the estimate of the cross-sections, we have used the radial wave functions from table~\ref{tab:fitresultall}. 
  In addition, the cross-section predictions will depend on $\alpha_s(\mu)$ and $m_c$. We have obtained the predictions using two different renormalization scales $\mu =2 m_c$, which will be scheme dependent, and $\mu = \sqrt{s}/2$ where $s$ is the square of the centre of mass energy for the productions of the charmonium states. We have taken the values of $\alpha_s(2 m_c)$ and $m_c$ from eq. \ref{eq:massavgscheme}, which are the averages over the three schemes and include the error due to scheme dependence. Using these inputs the predicted values of the respective cross sections are shown in table \ref{tab:epemdoublechar}. Note that given the errors our estimated values of $\sigma(e^+e^-\to J/\psi \eta_c)$ in both the renormalization scales are fully consistent with the current lower limit from BaBar and Belle. 

\subsection{Z boson decays into charmonium}

Using our results for the radial wave functions we have updated the rates of various $Z$ decays involving charmonium final states. These include radiative decays including $J/\psi$ or $\eta_c$ in the final state. Equally important channels are the exclusive $Z$ boson decays into charmonium-charmonium final states. We have updated the decay rates calculated at the leading order in NRQCD~\cite{Likhoded:2017jmx, Luchinsky:2017jab}. The corresponding expressions are given by  
\begin{eqnarray}
&\Gamma(Z\rightarrow\gamma\eta_{c})=\frac{2\alpha c_{V}^{2} e_{c}^{2} f_{\eta_{c}}^{2}}{3 M_{Z}}(1-4 r^{2}) \\ 
&\Gamma(Z\rightarrow\gamma J/\psi)=\frac{2\alpha c_{A}^{2} e_{c}^{2} f_{J/\psi}^{2}}{3 M_{Z}}(1-16 r^{4})
\end{eqnarray}
and 
\begin{eqnarray}
\Gamma(Z\to \eta_{c}  J/\psi) &=&\frac{8192 \pi  \alpha_s^2 \beta ^3 c_V^2  f_{\eta_c}^2 f_{J/\psi}^{2} r^2}{243 M_Z^3}\\ 
\Gamma(Z\to J/\psi J/\psi) &=&\frac{1024 \pi  \alpha_s^2 \beta ^5 c_A^2 f_{J/\psi}^{4} r^2}{243 M_Z^3}.
\end{eqnarray}
where $r=\frac{m_{c}}{M_{Z}}$. 
\begin{table}[ht!]
	\begin{center}
		\begin{tabular}{|*{2}{c|}}
			\hline
			$\textbf{Decay Modes}$ & $\textbf{Our Predictions}$ \\
			\hline
			$10^{12} \times\mathcal{B}(Z\to\eta _c+ J/\psi)$~\cite{Likhoded:2017jmx}&  $\text{0.0102(7)}$ \\
			$10^{12} \times\mathcal{B}(Z\to J/\psi$+$J/\psi)$~\cite{Luchinsky:2017jab}&  $\text{0.0101(5)}$  \\
			$10^{8} \times\mathcal{B}(\text{$Z \to \eta _c + \gamma$})$~\cite{Luchinsky:2017jab}&  $\text{0.545(35)}$  \\
			$10^{8} \times\mathcal{B}(\text{$Z \to J/\psi + \gamma $})$ ~\cite{Luchinsky:2017jab}& $\text{4.31(14)}$  \\
			\hline
		\end{tabular}
		\caption{Cross section estimates for double charmonium and radiative decays of the Z meson are obtained using the decay constant value from equation \ref{eq:avgdc}.}
		\label{tab:zdecays}
	\end{center}
\end{table}
From the decay rates $\Gamma(J/\psi \to e^+e^-)$ and $\Gamma(\eta_c\to \gamma\gamma)$ we obtain the following values of the respective decay constants
\begin{equation}
f^{\Gamma}_{J/\psi} = (349 \pm 001)~MeV, \ \  f^{\Gamma}_{\eta_c} = (292\pm 011)~MeV.
\end{equation}
The average values of these decay constants and the lattice predicted values shown in table \ref{tab:inputlattQCDSR} are given by	
\begin{equation}\label{eq:avgdc}
f^{avg.}_{J/\psi} = (377\pm 006)~~ \text{MeV}, \ \ f^{avg.}_{\eta_c} = (343 \pm 011)~\text{MeV}. 
\end{equation}

We provide numerical predictions for the BRs of Z boson decays to di-charmonium final states and radiative decays of Z boson into charmonium final states, which are presented in table~\ref{tab:zdecays}. Also, we provide the corresponding 1$\sigma$ CI. In this analysis, we have extracted the decay constants from a simultaneous analysis of the decay rates and the inputs on the decay constants from lattice given by eq.~\ref{eq:avgdc}. The earlier numerical analysis \cite{Likhoded:2017jmx, Luchinsky:2017jab,Gao:2022mwa,Li:2023tzx} mostly used model-dependent inputs on the radial wave functions or the inputs on $\Gamma(J/\psi \to e^+e^-)$ to extract $f_{J/\psi}$.

\section{Summary}	

This paper discusses the semileptonic and non-leptonic decays of the $B_c$ meson to S-wave charmonia, such as the $J/\psi$ and $\eta_c$ mesons. In a part of this analysis, using lattice inputs on the form factors and HQSS, we have extracted the shapes of the form factors $f_{+,0}^{B_c\to \eta_c}(q^2)$. In this process, we have also extracted the Isgur-Wise function defined at the zero recoil ($q^2_{max}$) in $B_c \to J/\psi (\eta_c)$ transitions and the parameters related to the symmetry-breaking effects at the non-zero recoils. Using these form factors we have predicted the $q^2$ integrated branching fractions $\mathcal{B}(B_c^- \to \eta_c\tau^- (\mu^-) \bar{\nu})$ and the LFU ratio of the rates $R(\eta_c)=0.310 \pm 0.042$ in the SM.

The $B_c\to J/\psi$ and $B_c\to \eta_c$ form factors, the decay constants $f_{J/\psi}$, $f_{\eta_c}$ and $f_{B_c}$ were calculated in the NRQCD effective theory. In the NRQCD, these form factors and the decay constants are sensitive to the radial wave functions $\psi_{B_c}^R(0)$, $\psi_{J/\psi}^R(0)$ and $\psi_{\eta_c}^R(0)$. In another part of our analysis, we have utilised the lattice inputs on $B_c\to J/\psi$ form factors (at the zero recoils), $B_c\to \eta_c$ form factors, the decay constants $f_{J/\psi}$, $f_{\eta_c}$ and $f_{B_c}$, and the measured charmonium rates $\mathcal{B}(J/\psi \to e^+e^-)$ and $\mathcal{B}(\eta_c \to \gamma \gamma)$ and extracted these radial wave functions. We have considered the NRQCD form factors up to the currently available order ``LO + NLO + RC". Furthermore, we have parametrised the uncomputed corrections and have treated them as nuisance parameters in the fit.
Using the extracted value for $\psi_{B_c}^R(0)$, $\psi_{J/\psi}^R(0)$ and $\psi_{\eta_c}^R(0)$ and the lattice inputs, we have predicted the branching fractions for the $B_c^- \to \eta_c \text{M}
(\text{M}= \pi^-, K^-,\rho^-, K^{*-})$ and $B_c^- \to J/\psi \text{M}
(\text{M}= \pi^-, K^-)$ decays, respectively. This, so far, is the first analysis where lattice inputs have been used in the computation of these predictions. As a final step, we have updated the numerical estimates of the cross sections $\sigma(e^+e^- \to J/\psi \eta_c, \eta_c\gamma)$ and predicted the branching fractions of $Z$ boson decays to either $J/\psi$ or $\eta_c$ or to both using our results on $\psi_{J/\psi}^R(0)$ and $\psi_{\eta_c}^R(0)$, respectively.

\section*{Acknowledgements}
A.B. received financial support from Spanish Ministry of Science, Innovation and Universities (project PID2020-112965GB-I00) and from the Research Grant Agency of the Government of Catalonia (project SGR 1069). S.S. acknowledges the Council of Scientific and Industrial Research (CSIR), Govt. of India for JRF fellowship grant with File No. 09/731(0173)/2019-EMR-I.

\section*{Appendix}
In this appendix, we present some supplemental information relevant to our analysis. 

\paragraph{\underline{Inputs relevant to HQSS relations}:}
We begin by presenting the definitions for $r$, $c_i$'s, $\theta$, $\omega$ relevant for the HQSS relations given in eq. \ref{eqn:FFrelns}. The detailed information can be seen in \cite{Cohen:2019zev} and references therein.
\begin{align}\label{eq:hqsscoeff}
 c^P_1&=1+\theta-\omega-\frac{\omega}{2}(1+\theta)(w-1) \, ,
\nonumber\\
 c^P_2&=1-\theta+\omega-\frac{\theta}{2}(1+\omega)(w-1) \, ,
\nonumber\\
 c_V&=-1-\theta-\omega \, , \nonumber\\
 c_\epsilon&=2-\frac{\omega+\theta+2\theta\omega-2}{2}(w-1) \, ,
\nonumber\\
 c_1&=\frac{(3+2\theta)\omega}{2} \, , \nonumber\\
 c_2&=-1 -\frac{\omega}{2} -\theta(1+\omega) \, ,
\end{align}
with $\theta \equiv \frac{m_{\bar{c}}}{2m_b}$, $\omega$ $\equiv$ $\frac{m_{\bar{c}}}{2 m_c}$ 
and $r=\frac{m}{m_{B_c}}$( where $m=m_{J/\psi}, m_{\eta_c}$ accordingly).

\paragraph{\underline{Lattice data points and HQSS parameters}:}

\begin{table}[t!]
\renewcommand*{\arraystretch}{1.4}
	\resizebox{1.\textwidth}{!}{
		\begin{tabular}{|c|cccccccccccc|}
			\hline
			\text{\bf Form factor Values} &\text{$\bf V(4)$}& \text{$\bf A_1(4)$}& \text{$\bf A_2(4)$}& \text{$\bf A_0(4)$}& \text{$\bf V(8)$}& \text{$\bf A_1(8)$}& \text{$\bf A_2(8)$}& \text{$\bf A_0(8)$}& \text{$\bf V(q^2_{\text{max}})$}& \text{$\bf A_1(q^2_{\text{max}})$}& \text{$\bf A_2(q^2_{\text{max}})$}& \text{$\bf A_0(q^2_{\text{max}})$}\\
			\text{\bf at diff. $\bf q^2$~(\textbf{in $GeV^2$})}&&&&&&&&&&&&\\
			\hline
			\text{\bf Input data} &\text{0.982(66)}&\text{0.563(25)}&\text{0.522(83)}&\text{0.639(30)}& \text{1.327(73)}&\text{0.705(25)}&\text{0.66(11)}&\text{0.854(33)}&\text{1.559(81)}& \text{0.801(27)}&\text{0.74(13)}&\text{0.996(37)}\\
			\hline\hline
			\text{\bf correlations}&1. & 0.036 & 0.006 & 0.034 & 0.941 & 0.037 & 0.005 & 0.032 & 0.863 & 0.035 & 0.005 & 0.028 \\
			&.& 1. & 0.63 & 0.463 & 0.036 & 0.923 & 0.506 & 0.456 & 0.034 & 0.825 & 0.414 & 0.427 \\
			&.&.& 1. & -0.356 & 0.007 & 0.463 & 0.867 & -0.296 & 0.007 & 0.335 & 0.742 & -0.246 \\
			&.&.&.& 1. & 0.032 & 0.54 & -0.296 & 0.929 & 0.03 & 0.55 & -0.245 & 0.836 \\
			&.&.&.&.& 1. & 0.039 & 0.007 & 0.032 & 0.981 & 0.037 & 0.007 & 0.03 \\
			&.&.&.&.&.& 1. & 0.444 & 0.54 & 0.038 & 0.972 & 0.401 & 0.504 \\
			&.&.&.&.&.&.& 1. & -0.334 & 0.007 & 0.335 & 0.977 & -0.326 \\
			&.&.&.&.&.&.&.& 1. & 0.031 & 0.561 & -0.324 & 0.977 \\
			&.&.&.&.&.&.&.&.& 1. & 0.037 & 0.008 & 0.029 \\
			&.&.&.&.&.&.&.&.&.& 1. & 0.311 & 0.53 \\
			&.&.&.&.&.&.&.&.&.&.& 1. & -0.336 \\
			&.&.&.&.&.&.&.&.&.&.&.& 1. \\
			\hline
		\end{tabular}
	}
	\caption{Pseudo data points and their correlations for the form factors in $B_c\to J/\psi$ decays at different $q^2$ values extracted from the results given in \cite{Harrison:2020gvo}.}
	\label{tab:corrBcjinputs}
\end{table}
\begin{table}[t]
\begin{center}
\begin{tabular}{|c|ccccc|}
\hline
&$\Delta_{V}(1)$&$\Delta_{A_1}(1)$&$\Delta_{A_0}(1)$&$\Delta(1)^{\prime}$&$\Delta(1)^{\prime \prime}$\\
   \hline
$\Delta_{V}(1)$&1. & 0.151  & 0.086 & -0.327 & 0.135  \\
$\Delta_{A_1}(1)$&.& 1. & 0.561 & -0.336 & 0.161   \\
    $\Delta_{A_0}(1)$&. & . & 1. & -0.358 & 0.203   \\
$\Delta(1)^{\prime}$&. & . & . & 1. & -0.721  \\
$\Delta(1)^{\prime \prime}$&. & . & . & . & 1.  \\
\hline
\end{tabular}
\caption{Correlations matrix between the HQSS parameters. Corresponding fit results are given in table \ref{tab:HQSSfitparam} (third column).}
\label{tab:corrHQSSparm2}
\end{center}
\end{table}
\begin{table}[t!]
\centering
\begin{tabular}{|c|cccccc|}
\hline
&$f_+(6)$&$f_+(8)$&$f_+(q^2_{\text{max}})$&$f_0(6)$&$f_0(8)$&$f_0(q^2_{\text{max}})$\\ \hline
$f_+(6)$&1. & 1. & 0.999 & 0. & 0. & 0. \\
$f_+(8)$& .& 1. & 1. & 0. & 0. & 0. \\
$f_+(q^2_{\text{max}})$& . & . & 1. & 0. & 0. & 0. \\
$f_0(6)$& . & . & . & 1. & 1. & 0.999 \\
$f_0(8)$& . & . & . & . & 1. & 1. \\
$f_0(q^2_{\text{max}})$& . & . & . & . & . & 1. \\
\hline
\end{tabular}
\caption{Correlations matrix of the obtained for the synthetic data points of $f_{+,0}(q^2)$ at a few specific values of $q^2$. The synthetic data points are shown in table \ref{tab:syndatfpf0}.}
\label{tab:corrfp0}
\end{table}

\begin{table}[h!]
\centering
	\begin{tabular}{|c|cccc|}
		\hline
		&$f_0^{a_0}$&$f_0^{a_1}$&$f_+^{a_0}$&$f_+^{a_1}$\\
		\hline
		$f_0^{a_0}$&1. & 0.979 & 0.998 & 0.945 \\
		$f_0^{a_1}$&. & 1. & 0.986 & 0.934 \\
		$f_+^{a_0}$&.& . & 1. & 0.932 \\
		$f_+^{a_1}$&. & .& . & 1. \\
		\hline
	\end{tabular}
\caption{Correlations matrix between the BCL coefficients. Corresponding fit results are given in table \ref{tab:BCLcoeffhqss}.}
\label{tab:corrBCLcoeff}
\end{table}
The synthetic lattice data points for $B_c\to J/\psi$ form factors used in our analysis alongside their correlations are given in table \ref{tab:corrBcjinputs}. We have created the pseudo data points at $q^2=4$, $8$ and $(m_{B_c}-m_{J/\psi})^2$ and estimated the correlations between them. Using these synthetic data points, we have fitted the HQSS parameters; we have shown the fit results in table \ref{tab:HQSSfitparam} and the corresponding correlations are shown in table \ref{tab:corrHQSSparm2}. Using these results and the HQSS symmetry, we have obtained the synthetic data points for the $B_c\to \eta_c$ form factors, which we have presented in table \ref{tab:syndatfpf0}. The respective correlation matrix is shown in table \ref{tab:corrfp0}. Using these synthetic data points on $B_c \to \eta_c$ form factors, we have extracted the BCL coefficients of the $z$ expansion of the form factors $f_{+,0}(q^2)$. We have presented the best-fit values of the coefficients in table \ref{tab:BCLcoeffhqss} and the corresponding correlation matrix in table \ref{tab:corrBCLcoeff}.

\paragraph{\underline{Available corrections in $B_c\to J/\psi$ form factors and $f_{B_c}$ in NRQCD}:}
Including the known estimates for the NLO-QCD and the LO relativistic corrections, the decay constants $f_{B_c}$ and the $B_c \to J/\psi$ form factors can be expressed as   
\begin{equation}\label{eq:nloRCdecay}
f_{B_c}= f^{LO}_{B_c} \big[ 1 - 0.094(15)|_{\mathcal{O}(\alpha_s)} - 0.034(7)|_{\mathcal{O}(v^2)}\big]
\end{equation}
and 
\begin{align}\label{eq:nloRCavgscheme}
A_0(0) &=A_0^{LO}(0)\big[1 + 0.306(48)|_{\mathcal{O}(\alpha_s)}+0.365(168)|_{\mathcal{O}(v^2)}\big],\nonumber \\
A_1(0) &=A_1^{LO}(0)\big[1 + 0.356(51)|_{\mathcal{O}(\alpha_s)} +0.392(160)|_{\mathcal{O}(v^2)}\big],\nonumber \\
A_2(0) &=A_2^{LO}(0)\big[1 + 0.356(51)|_{\mathcal{O}(\alpha_s)}+0.434(160)|_{\mathcal{O}(v^2)}\big],\nonumber \\
V(0) &=V^{LO}(0)\big[1 + 0.356(51)|_{\mathcal{O}(\alpha_s)} + 0.443(162)|_{\mathcal{O}(v^2)}\big],
\end{align}
respectively. The numerical estimates in these equations are based on the inputs we provide in eq. \ref{eq:massavgscheme}, and the errors are due to the scheme dependences of $m_b$, $m_c$ and $\alpha_s$. The higher order corrections in the NRCQD form factors are large as compared to those in the decay constants, hence, impact more on the extracted values of $\psi^R_{B_c}(0)$.
\begin{table}[t!]
\begin{center}
\scriptsize
\renewcommand*{\arraystretch}{1.6}
\resizebox{1.03\textwidth}{!}{
\begin{tabular}{c|cccccc}
\hline\hline
\multicolumn{1}{c|}{}&  ~~~~$\psi^R_{B_c}$(0)~~~~ &~~~~\text{$\psi^R_{J/\psi}(0)$}~~& ~~~~\text{$\psi^R_{\eta _c}(0)$}~~~~&~~\text{$\delta_{f_{B_c}}$}~~&~~\text{$\delta_{J\psi ee}$}~~&~~\textbf{$\delta_{\eta_c \gamma \gamma}$} \\
\hline
\multirow{1}{*}{\bf (I) LO}
& $\text{0.96(13)}$ &  $\text{0.836(39)}$  &  $\text{1.029(26)}$ &$\text{0.27(15)}$&$\text{0.002(58)}$&$\text{-0.101(26)}$  \\
\hline
\multirow{1}{*}{\bf (II) LO+NLO}
&  $\text{0.68(11)}$&  $\text{0.836(39)}$ & $\text{1.028(26)}$ &$\text{0.75(25)}$&$\text{0.003(58)}$&$\text{-0.100(26)}$\\
\hline
\multirow{1}{*}{\bf (III) LO+NLO+RC}
&  $\text{0.61(9)}$  &  $\text{0.829(40)}$  &$\text{1.029(26)}$
&$\text{0.92(26)}$&$\text{0.009(58)}$&$\text{-0.100(26)}$\\
\hline\hline
\end{tabular}
}
\caption{Fit results for radial wave functions corresponding to the $B_c$, $J/\psi$ and $\eta_c$ mesons. We use $B_c \to J/\psi$ and $B_c \to \eta_c$ form factors at $q^2=0$ at different orders in NRQCD and the observables BR($J/\psi\to e^+ e^-$) and BR($\eta_{c}\to \gamma\gamma$) and decay constants for $f_{B_c}, f_{J/\psi}, f_{\eta_c}$ as inputs to the fit. We have also considered additional errors ($\delta_i$) due to missing higher-order corrections as nuisance parameters.}
\label{tab:fitresultsNRQCDWF}
\end{center}
\end{table}

In table \ref{tab:fitresultsNRQCDWF}, we have presented the result of the fits where the NRQCD $B_c \to J/\psi (\eta_c)$ form factors are considered, respectively, at LO, (LO + NLO) and (LO + NLO +RC). For the decay constants, the available higher-order perturbative and relativistic corrections are included. Also, the uncomputed corrections are parametrised as nuisance parameters ($\delta_i$s). For the details, see the main text. Note that the extracted values of $\psi^R_{B_c}(0)$ gradually reduce with the inclusion of higher order corrections to $B_c\to J/\psi$ form factors. At LO, (LO+NLO), and (LO+NLO+RC) in all the three schemes, the extracted errors in $\psi^R_{B_c}(0)$ are the same, which is about 15\%, the reason being that the primary sources of errors are the lattice inputs on the form factors. We can understand the impact of contributions at NLO and RC from the changes in the best-fit values of $\psi^R_{B_c}(0)$ in the respective fits from that of LO results. The percentage changes in $\psi^R_{B_c}(0)$ precisely reflect the size of the contributions at the respective order (NLO or RC) with respect to the LO contributions. The reduction in the values of $\psi^R_{B_c}(0)$ is because the higher order corrections in all these form factors appear with a plus sign. There are, hence, no relative numerical cancellations between different orders.
\bibliographystyle{JHEP}
\bibliography{Bcanylz.bib} 

\end{document}